# Isoleucine gate blocks K+ conduction in C-type inactivation

Werner Treptow,[1,2*] Yichen Liu,[3] Carlos A. Z. Bassetto Jr,[2] Bernardo I. Pinto,[2] João Alves,[1] Christophe Chipot[2,4,5], Francisco Bezanilla[2,6] and Benoît Roux[2*]

[1]Laboratório de Biologia Teórica e Computacional (LBTC), Universidade de Brasília, DF, Brasil, CEP 70904-970
[2]Department of Biochemistry and Molecular Biology, The University of Chicago, Chicago, Illinois 60637, United States
[3]Department of Neurobiology, The University of Chicago, Chicago, Illinois 60637, United States
[4]Laboratoire International Associé Centre National de la Recherche Scientifique et University of Illinois at Urbana–Champaign, Unité Mixte de Recherche No. 7019, Université de Lorraine, Université de Lorraine, B.P. 70239, 54506 Vandœuvre-lès-Nancy cedex, France
[5]NIH Center for Macromolecular Modeling and Bioinformatics, Beckman Institute for Advanced Science and Technology, and Department of Physics, University of Illinois at Urbana–Champaign, Urbana, Illinois 61801, United States
[6]Centro Interdisciplinario de Neurociencia de Valparaíso, Facultad de Ciencias, Universidad de Valparaíso, Valparaíso, Chile

**WT, YL and CB contributed equally to the work**

**[*]To whom correspondence may be addressed. E-mail.** treptow@unb.br, roux@uchicago.edu

**ABSTRACT:** Many voltage-gated potassium (Kv) channels display a time-dependent phenomenon called C-type inactivation, whereby prolonged activation by voltage leads to the inhibition of ionic conduction, a process that involves a conformational change at the selectivity filter toward a non-conductive state. Recently, a high-resolution structure of a strongly inactivating triple-mutant channel kv1.2-kv2.1-3m revealed a novel conformation of the selectivity filter that is dilated at its outer end, distinct from the well-characterized conductive state. While the experimental structure was interpreted as the elusive non-conductive state, molecular dynamics simulations and electrophysiology measurements demonstrate that the dilated filter of kv1.2-kv2.1-3m, however, is conductive and, as such, cannot completely account for the inactivation of the channel observed in functional experiments. An additional conformational change implicating isoleucine residues at position 398 along the pore lining segment S6 is required to effectively block ion conduction. It is shown that the I398 residues from the four subunits act as a state-dependent hydrophobic gate located immediately beneath the selectivity filter. As a critical piece of the C-type inactivation machinery, this structural feature is the potential target of a broad class of QA blockers and negatively charged activators thus opening new research directions towards the development of drugs that specifically modulate gating-states of Kv channels.

## INTRODUCTION

The selectivity filter of voltage-gated potassium (Kv) channels is a specialized molecular structure, responsible for the fast and selective conduction of potassium ($K^+$) over other ionic species. It is well-established that the selectivity filter of Kv channels can exist in both conductive and non-conductive states (Fig. 1). Known as the process of (slow) C-type inactivation (*1*, *2*), the conductive to non-conductive conformational transition is of great physiological importance as it contributes to fine-tune long-term activity of Kv channels. The canonical conductive conformational state of the filter was first revealed with the crystallographic structure of the prototypical bacterial $K^+$ channel KcsA at high resolution (*e.g.* pdb id 1BL8 (*3*) or 1K4C (*4*)). MD studies confirm that ion conduction along this filter conformation is possible and unopposed by large free energy barriers (*5*,

*6*). Since, additional structures have been resolved for other Kv channels of the (eukaryotic) *Shaker* family, broadening our knowledge of the conductive state of the filter (*7*). Recently, high-resolution structures of Kv channels revealed a novel conformation of the selectivity filter that is partially dilated at its outer end and constricted near its internal face (*8–10*). Because a number of mutations known to strongly enhance the process of C-type inactivation were introduced in the construct used in structural determination (*11*), this "dilated" conformation has been interpreted as the non-conductive conformational state of the selectivity filter, accounting for the phenomenon of C-type inactivation. Reinforcing that notion, the most recent structure of the wild-type *Shaker* B channel displays the same dilated conformation of the filter at low-concentration of external potassium (*12*).

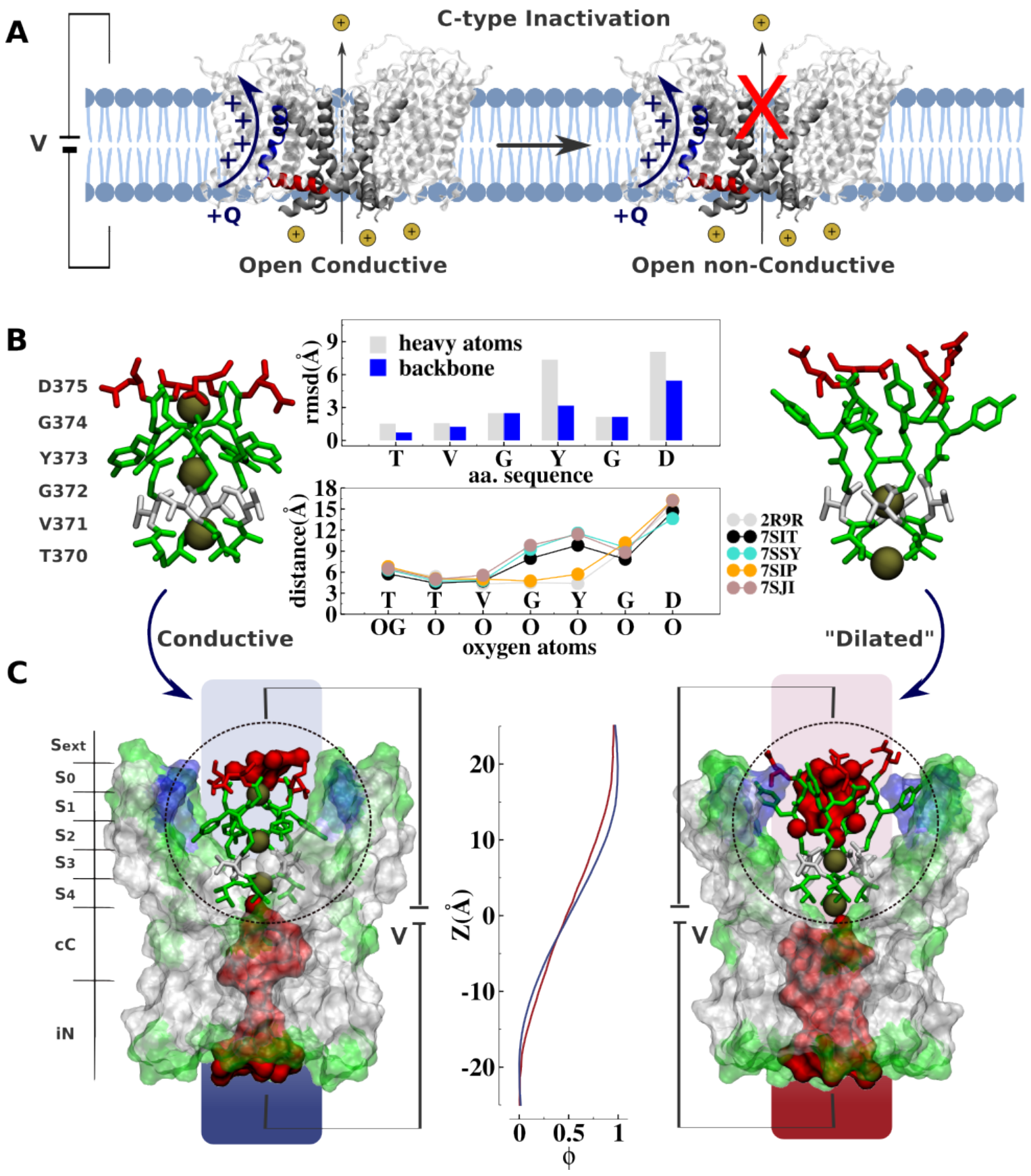

**Fig. 1. Comparative analysis of Kv channel structures.** (**A**) Schematic representation of a voltage-gated $K^+$ channel undergoing C-type inactivation, whereby prolonged activation by an external voltage V leads to blockage of ionic conduction across the selectivity filter of the open channel. The voltage-sensor positively charged S4 helix (blue), the S4S5 linker (red) and main-pore S6 helix (silver) are highlighted. (**B**) Structural models of the selectivity filter in the conductive and dilated conformations. The conductive and dilated conformations derive respectively from the high-resolution x-ray structures of the wild-type (*7*) and triple-mutant (*9*) kv1.2-kv2.1 channel (PDB codes 2R9R and 7SIT). Major structural deviations (rmsd) between the selectivity filter conformations are primarily accounted by side-chain rearrangements of Y373 and D375. Despite the side-chain rearrangements of Y373 and D375, the profile of oxygen-oxygen distances between opposing subunits of the selectivity filter indicates that the geometry of sites $S_4$ and $S_3$ in the dilated conformation closely resembles that of the conductive state. For comparison purposes, the profile of oxygen-oxygen distances is also shown along the selectivity filter of the experimental structures of the conductive state of *Shaker* B (*8*) (PDB code 7SIP) and the dilated conformations of *Shaker*-W434F (*8*) (PDB code 7SJI) and Kv1.3 (*13*) (PDB code 7SJ1). (**C**) Electrostatic properties of the conductive and dilated conformations of the selectivity filter. Shown are molecular representations of the main-pore S6 segments of the channel, highlighting the permeation pathway along the intracellular entrance (iN), central cavity (cC) and selectivity-filter sites ($S_4$, $S_3$, $S_2$, $S_1$, $S_0$, $S_{ext}$). Hydrated cavities (red) give ionic access to the selectivity filter from the intracellular and extracellular milieu. The dielectric morphology of the protein and waters accounts for a significant voltage drop (ϕ) across the selectivity filter. The voltage-drop profile along the permeation pathway was computed as described elsewhere (*14*), following the charge-imbalance protocol which is a variant of the linear field method (*15*).

The “dilated” and “conductive” filter conformations differ markedly. The all-atom root-mean-square (RMS) deviations of the selectivity filter between the dilated and the conductive conformation is about 3.0 Å, arising mostly from rearrangements of the side chain of the highly conserved tyrosine and aspartic acid along the signature of the selectivity filter TVGYGD (Fig. 1). Yet, despite the considerable conformational differences, a simple inspection shows no apparent physical barrier to ion transport along the permeation pathway of the “dilated” structure, challenging its postulated non-conductivity (inactive). The selectivity filter is accessible to the intracellular solution via a large open and hydrated vestibular cavity. Incoming ions can bind to sites $S_4$ and $S_3$ and translocate to the extracellular side via a a hydrated crevice corresponding to the widened sites $S_2$, $S_1$ and $S_0$. The open and fully hydrated vestibular cavity is clearly conductive, which disagrees with previous cysteine modification and blocker-protection assays indicating that the cavity of the channel is changed in the slow-inactivated state compared with the open state (*16*). Besides, the voltage drop across the selectivity filter through the conductive and dilated conformations is similar (Fig. 1C). In contrast to classic chemical and peptide blockers or the constricted filter associated with the inactivation of the KcsA bacterial channel that typically inhibit conduction by providing a physical barrier along the permeation pathway (*2*, *10*, *17–19*), the dilated conformation does not appear to impose any structural impediment to ionic conduction. These observations beg for a quantitative assessment of the ion conduction properties of the dilated conformation, and the relation of the latter to the functional C-type inactivated state identified in experiments.

## RESULTS

To address these questions, detailed all-atom molecular dynamics (MD) simulations were carried out on the basis of the high-resolution X-ray structure of the kv1.2-kv2.1-3m chimera channel containing the triple mutation W362F, S367T and V377T (*9*). To circumvent the uncertainties inherent with MD and achieve robust conclusions, three different force fields, AMBER (*20*, *21*), CHARMM36m (*22*) and a modified CHARMM36m-NBFIX were considered (Methods, Tables S1 and S2). In all the simulations, the channel was embedded in a fully hydrated phospholipid bilayer at 150mM KCl and simulated with an applied transmembrane of +200mV.

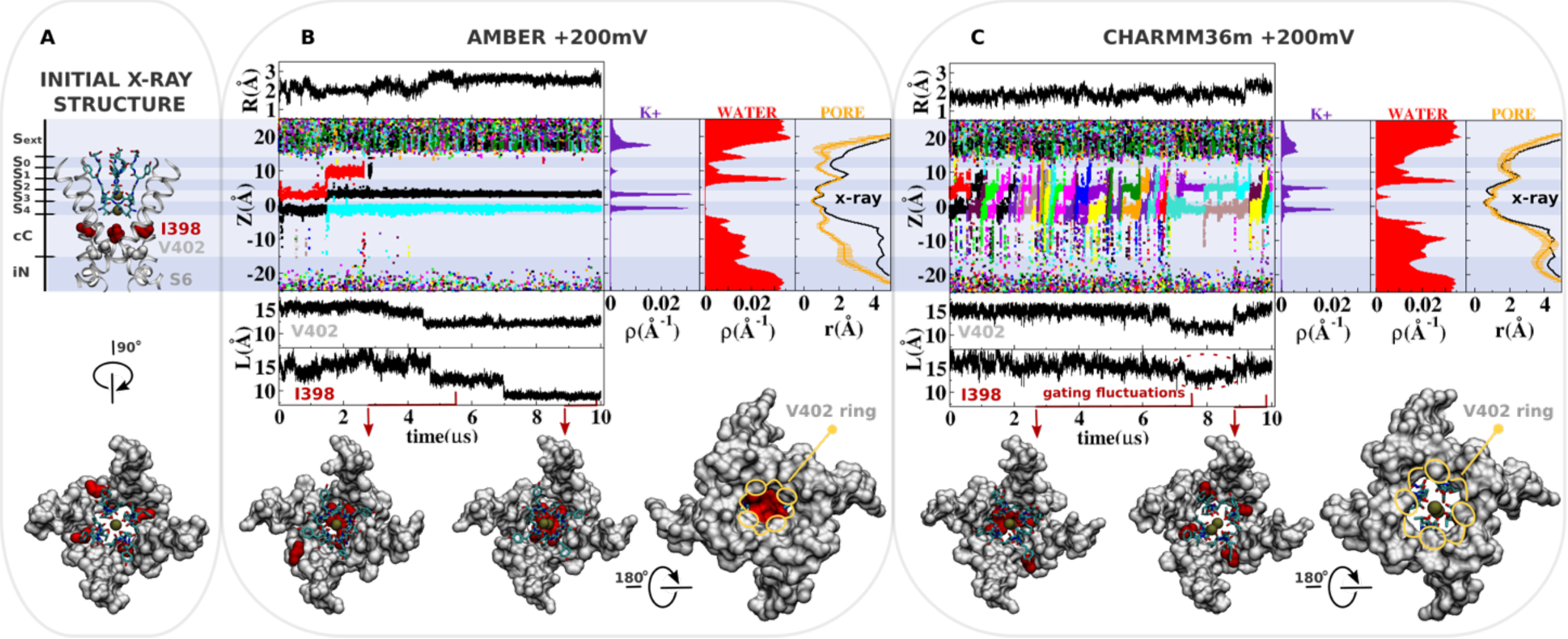

**Fig. 2. MD simulation of kv1.2-kv2.1-3m at +200mV.** (**A**) Molecular representation of the main pore of the channel, highlighting the initial configuration of the selectivity filter, I398 (red) and V402 (light gray). (**B** and **C**) Analysis of AMBER and CHARMM36m trajectories. Shown is the structural deviation of the selectivity filter (R), the trajectory of $K^+$ ions along the permeation pathway (Z) and the intersubunit $C_\beta$-$C_\beta$ separation distance of I398 and V402 (L) as a function of simulation time. Inset shows instantaneous configurations of I398 (red arrows). Time averages are the linear density of $K^+$ ions, the linear density of water oxygen and the pore radius profile (r) along the permeation pathway Z.

The kv1.2-kv2.1-3m channel with the dilated selectivity filter appears remarkably stable during a 10μs simulation using the AMBER force field (Fig. 2). The structural RMS deviation of the selectivity filter relative to the initial X-ray structure (*9*) is less than 2.8 Å, indicative of structural stability (*cf.* Fig. 2B, structural deviation R). Simulation of kv1.2-kv2.1-3m with the CHARMM36m force field also illustrates the stability of the dilated conformation over the microsecond timescale, with structural deviation of the selectivity filter less than 2.5 Å (*cf.* Fig. 2C, structural deviation R). The average density of the $K^+$ ions along the permeation pathway from the AMBER trajectory is consistent with a predominant occupancy of the sites $S_4$, $S_3$ and $S_{ext}$. Not resolved in the x-ray structure of kv1.2-kv2.1-3m (*9*), the density at the external site $S_{ext}$ results from close interactions of $K^+$ and the carboxylate group of D375. The acidic side chain adopts a relaxed conformation throughout the simulation, in which the carboxylate moiety is fully exposed to the external solution in an orientation similar to that previously reported in the cryogenic electron microscopy (cryo-EM) structure of the homologous Kv1.3 channel with a dilated selectivity filter (*13*). According to a knock-on mechanism at the level of binding sites $S_4$ and $S_3$ (Fig. S1) (*12*), there is an early voltage-driven $K^+$ conduction event across the selectivity filter within the first ~2μs of the simulation (*cf.* Fig. 2B, trajectory of $K^+$ along the permeation pathway Z). Conduction increases the residence time of the ion in the dilated region of the selectivity filter and accounts for a minor density peak at the level of site $S_1$ (*cf.* Fig. 2B, linear density of $K^+$), which is consistent with the resolved electron density of the cation in the experimental structures (*8–10*, *12*). Ion access to the central cavity of the channel and their subsequent conduction across the selectivity filter completely cease after ~4μs of simulation time when the side chain of isoleucine at position 398 from all four S6 segments twist toward the central axis of the channel, dehydrating and blocking the permeation pathway immediately beneath the selectivity filter (*cf.* Fig. 2B, density of water along the pore axis). The $C_\beta$-$C_\beta$ distance between I398 in opposing subunits decreases over a period of 3-4μs, reaching a value of about 9Å—in stark contrast with the initial distance of 15Å in the X-ray structure (*9*) (*cf.* Fig. 2, separation distance L of I398). As a consequence, the pore radius becomes locally constricted ( ≤ 2Å), blocking ion conduction (*cf.* Fig. 2B, pore radius r) (*23*).

In contrast, ion conduction events are observed in the simulation with the CHARMM36m force field. This observation is in sharp contrast with the conclusion drawn from previous simulations of *Shaker* B also based on the CHARMM36m force field, where no ion conduction event was observed even with an applied membrane potential of +300mV (*8*, *12*). Importantly, dihedral restraints were applied in these simulations to preclude deviations from the X-ray structure, which presumably inhibited ion conduction through the dilated filter. Because the CHARMM36m force field favors the open configuration of the isoleucine gate, the simulation shows that the dilated conformation is highly conductive under membrane depolarization. The distribution of ions along the permeation pathway in the CHARMM36m simulation is distinct from that inferred from the AMBER simulations, as the increased mobility of $K^+$ in the selectivity filter accounts for a more pronounced reallocation of the ionic density from sites $S_4/S_3$ to site $S_2/S_1$.

To further clarify the ion conduction properties of the dilated filter, we simulate the channel with the AMBER force field in the presence of harmonic distance restraint to keep the isoleucine gate in the open configuration (Fig. S2). Compared to the unrestrained simulation with the AMBER force field, a larger number of ions access the central cavity of the channel, triggering two spontaneous conduction events across the selectivity filter in the early ~4μs of the simulation. The density profile of the $K^+$ ion along the channel axis is distinct from that of the unconstrained simulation. Partial reallocation of the ionic density from sites $S_4/S_3$ to sites $S_2/S_1$ reflects the increased mobility of $K^+$ in the selectivity filter, thereby corroborating the hypothesis that the dilated conformation of the selectivity filter is conductive when the isoleucine gate is open. This conclusion is further confirmed from additional simulations based on a modified CHARMM36m-NBFIX hybrid model in which ion-water, ion-protein and water-protein interactions in CHARMM36m are fine-tuned to mimic those in the AMBER force field (Fig. S2, Table S2). In the CHARMM36m-NBFIX simulation, the local concentration of ions in the central cavity of the channel is significantly smaller than that of the original trajectory. Pronounced density peaks at sites $S_4/S_3$ indicates that $K^+$ binds more strongly to the selectivity filter. The combined effect yields a conductive AMBER-like channel, characterized by fewer conduction events per simulation time.

These simulations strongly suggest that the filter in the dilated conformation can conduct $K^+$ ions, and that the conformational motion of I398 is necessary to truly block conduction. Thus, it is apparently the conformational change implicating the isoleucine gate that leads the channel toward a true non-conductive state. As a significant modification of the channel structure, the local rearrangement of I398 is coupled to motions of other regions of the main pore, including V402 at the highly conserved PVP motif (Fig. 2). Analysis indicates that CHARMM36m favors the open and fully hydrated state of the I398 gate. The average intersubunit $C_\beta$-$C_\beta$ distance of the isoleucine side chains (~16Å) is close to the reference value in the X-ray structure (~15Å) most of the simulation time. Gating fluctuations of I398 are, however, clearly observed in the late stages of the simulation and, correlate well with the reduction of ions in the central cavity of the channel and with the conduction across the selectivity filter. Particularly important, the CHARMM36m simulation adds support to the assumption that the dilated conformation of the selectivity filter is conductive and that closure of the isoleucine gate is required to shut down ion transport across the channel.

Despite intrinsic force-field differences with respect to channel conductivity, all three atomistic models support that the dilated conformation of the selectivity filter is, by itself, conductive and the isoleucine gate is required to effectively block $K^+$ current across the channel. According to the voltage-driven MD trajectories in which the isoleucine gate is open, the total number of conduction events across the dilated conformation of the selectivity filter over the total simulation time is 44(ions)/30μs (Table S3). Based on the number of crossing events per simulation time, the ionic current of a macroscopic population of ion channels in the open-gate dilated state is ~0.2pA at the TM voltage of +200mV (single-channel conductance of ~1.17pS). In all likelihood underestimation of the channel conductance as a consequence of the well-documented force-field limitations in reproducing the ionic current in Kv channels (*24*), the estimate of ~0.2pA is still orders of magnitude larger than the measured current in the triple-mutant channel upon C-type inactivation (v*ide infra*), and, therefore, the conductivity properties of the "dilated" conformation of the selectivity filter cannot explain alone the inactivation of kv1.2-kv2.1-3m under membrane depolarization. Consistent with single-channel measurements (*25*), the estimate of ~1.17pS is actually more comparable to simulation predictions of the single-channel conductance of the conductive selectivity filter *i.e.*, ~3.5pS (*12*). Beside the dilated conformation of the selectivity filter, the isoleucine gate then appears to be a critical molecular element of the channel machinery, largely implicated in C-type inactivation.

To validate and corroborate the key role of the isoleucine gate in C-type inactivation inferred from the MD simulations, electrophysiology experiments were carried out (Methods, Fig. 3). In stark contrast with recordings of the kv1.2-kv2.1 chimera channel (Fig. 3A), C-type inactivation is greatly enhanced in the triple-mutant channel kv1.2-kv2.1-3m, as is evidenced by the fast decay of the ionic current in the ms timescale and the appearance of a gating current (Fig. 3B and Fig. S3). However, substitution of the isoleucine by the polar amino-acid asparagine, with similar side-chain volume, restores the conductivity of the triple-mutant channel (Fig. 3C). While not disturbing the steady-state current-voltage relationship of the triple-mutant channel, mutation I398N drastically increases ionic conduction without any apparent time-dependent inactivation (Fig. 3D and E). Importantly, restoration of ion conduction with I398N is not caused by inadvertently stabilizing the selectivity filter in the conductive state, as demonstrated by pore-blocking toxin assays. Agitoxin-II (AgTxII), Dendrotoxin (DTX) and Charybdotoxin (CTX) are potent toxin blockers of Kv channels (*26*), binding to the outer mouth of the channel (*10*, *17*, *18*). Such pore-blocking toxins preferentially bind and occlude the conductive conformation of the selectivity filter of Kv channels, as indicated by previous functional studies (*27*) as well as implicit-solvent binding free-energy calculations (Methods, Fig. 3F, Fig. S4, Table S4). Whereas AgTxII blocks the ionic current across the conductive kv1.2-kv2.1 channel by binding to the outer mouth of the filter in the conductive conformation (Fig. 3G), it fails to bind and occlude the kv1.2-kv2.1-3m channel with its filter in the dilated conformation which then becomes conductive with the I398N mutation (Fig. 3I). Because the slow-inactivated state of kv1.2-kv2.1-3m is itself immune to toxin effects (Fig. 3H), ion conduction with the I398N mutation proves that the permeation pathway must be altered at some location along the permeation pathway other than the filter. Such structural modifications of the permeation pathway are specific of C-type inactivation since a double-mutant (S367T/V377T) of the chimera channel kv1.2-kv2.1-2m that does not inactivate is vulnerable to AgTxII, irrespective of the I398N mutation (Fig. S5). The critical C-type inactivation mutation W362F (*11*) is missing in the kv1.2-kv2.1-2m channel, hence its inability to slow-inactivate.

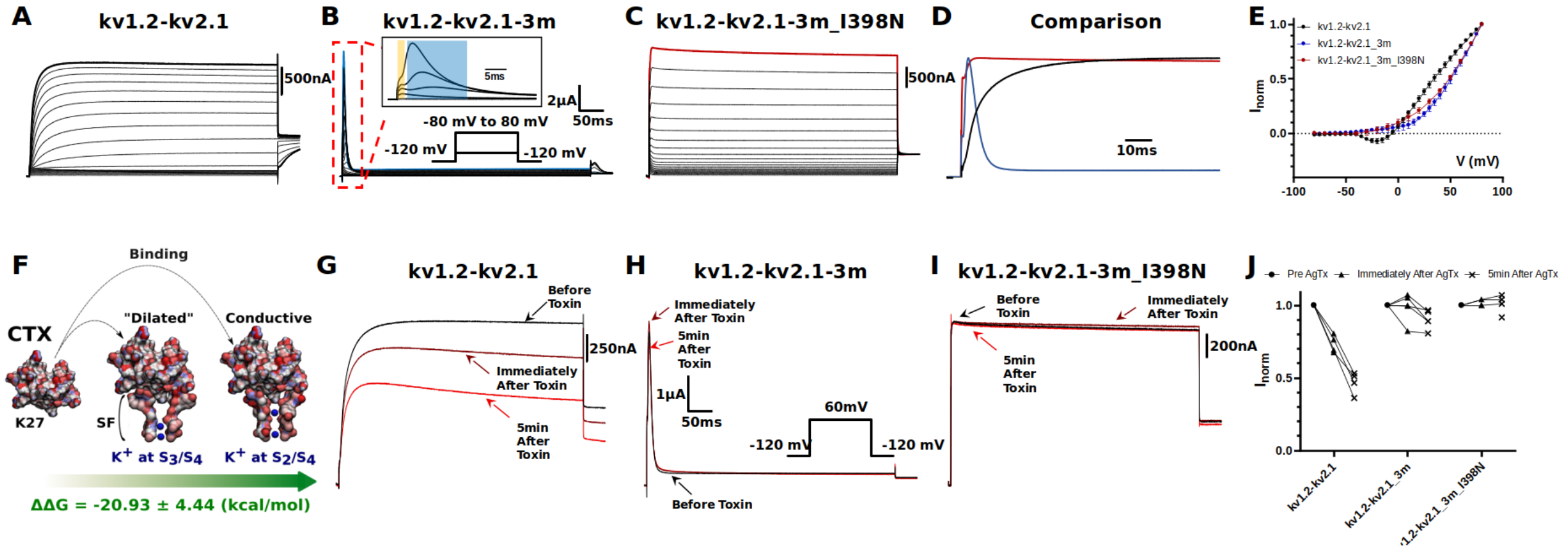

**Fig. 3. Electrophysiology measurements of I398N substitution.** (**A**, **B**, **C** and **D**) Macroscopic current recorded from: (**A**) kv1.2-kv2.1 chimera, (**B**) triple mutant kv1.2-kv2.1-3m, (**C**) triple mutant kv1.2-kv2.1-3m with I398N substitution and (**D**) their respective comparison at +80mV (the line colors correspond to the colors indicated in **A**, **B** and **C** of the three mutants at +80 mV). The triple mutation W362F, S367T and V377T in kv1.2-kv2.1-3m significantly speed up the inactivation process and the gating current could be seen simultaneously with ionic current (shown in inset, with gating current highlighted in yellow and ionic current highlighted in blue). Note the effect of the triple mutation cancel out with the I398N substitution. (**E**) Steady-state current-voltage relationship. (**F**) Net free-energy difference (ΔΔG) involved in the binding of the Charybdotoxin (CTX) to the conductive (*18*) and dilated conformations of the selectivity filter of the kv1.2-kv2.1 chimera channel. ΔΔG indicates a strong preference of CTX to the conductive conformation of the selectivity filter. The same binding preference is found between CTX and kv1.2-kv2.1-3m (Table S3). (**G**, **H**, **I** and **J**): Effects of Agitoxin-II, a more potent CTX analog (*26*), on kv1.2-kv2.1 chimera, triple mutant kv1.2-kv2.1-3m, triple mutant kv1.2-kv2.1-3m with I398N substitution. Clearly, Agitoxin binds and blocks the chimera channel while shows minimal influence on triple and triple_I398N, suggesting the selectivity filter in triple_I398N likely also adopts a dilated conformation.

Consistent with experiment, a polar amino acid allows ion conduction by keeping the permeation pathway constitutively hydrated and open. Additional AMBER and CHARMM36m simulations of the triple-mutant channel support this view by revealing that I398N prevents closure of the gate at position 398 (Table S1, Fig. S6). The asparagines in all four S6 segments twist their side chains toward the aqueous central cavity of the channel, causing a decrease in the average intersubunit $C_\beta$-$C_\beta$ distance. In contrast with I398, the open and hydrated configuration of the gate in the I398N mutant allows ion conduction across the channel as long as structural fluctuations of the PVP motif (V402) do not obstruct the permeation pathway. Clearly shown in both simulations, the coupled motions between N398 and V402 impact the local concentration of incoming ions, and the ensuing number of crossing events across the channel over the microsecond timescale.

## DISCUSSION

Taken together, the present structural and functional results demonstrate that the dilated conformation of the selectivity filter of kv1.2-kv2.1-3m is conductive and that the isoleucine gate is critical to block $K^+$ currents during C-type inactivation of the channel. Judged by the primary-sequence conservation of I398 (Fig. S7), the isoleucine gate seems to be relevant for potassium channels that undergo C-type inactivation in general, and, potentially, for other voltage-gated $Na^+$ channels possessing distinct selectivity filters (*28*). The action of the isoleucine gate in C-type inactivation is averted by the I398N mutation because the permeation pathway remains constitutively open and hydrated with the polar asparagine side. In the homologous *Shaker* B channel, the single I470C (*29*, *30*) and double T449V/I470C (*31*) mutations at the corresponding position also convert the slow inactivated state into a conductive state, thus eliminating the slow-inactivation phenotype under long depolarizations strongly corroborates our findings. Closing of the I398 gate in the dilated conformation of kv1.2-kv2.1-3m involves occlusion of the binding site of internally applied quaternary ammonium (QA) blockers (*32*), explaining the previously reported 20-fold decreased affinity of TEA for the inactivated state of *Shaker*-IR compared with that of the open state (*16*). On the other hand, the mechanism whereby I398N renders the gate constitutively open, not occluding the binding site of QA compounds, is also consistent with the demonstration that I470C in *Shaker*-IR morphs the channel that does not trap quaternary ammoniums into one that does (*29*). Across all these measured effects, the *modus operandi* of the isoleucine gate is expected to be coupled to the PVP motif, and, therefore, to reflect to some extent the conformational allostery between the selectivity filter and the bundle-crossing region previously reported for C-type inactivation (*16*, *33*, *34*). While these experimental results were indicative of the role of the isoleucine gate its mechanistic significance with regards to the non-conductive C-type inactivation seems to have been overlooked (*8*–*10*, *12*).

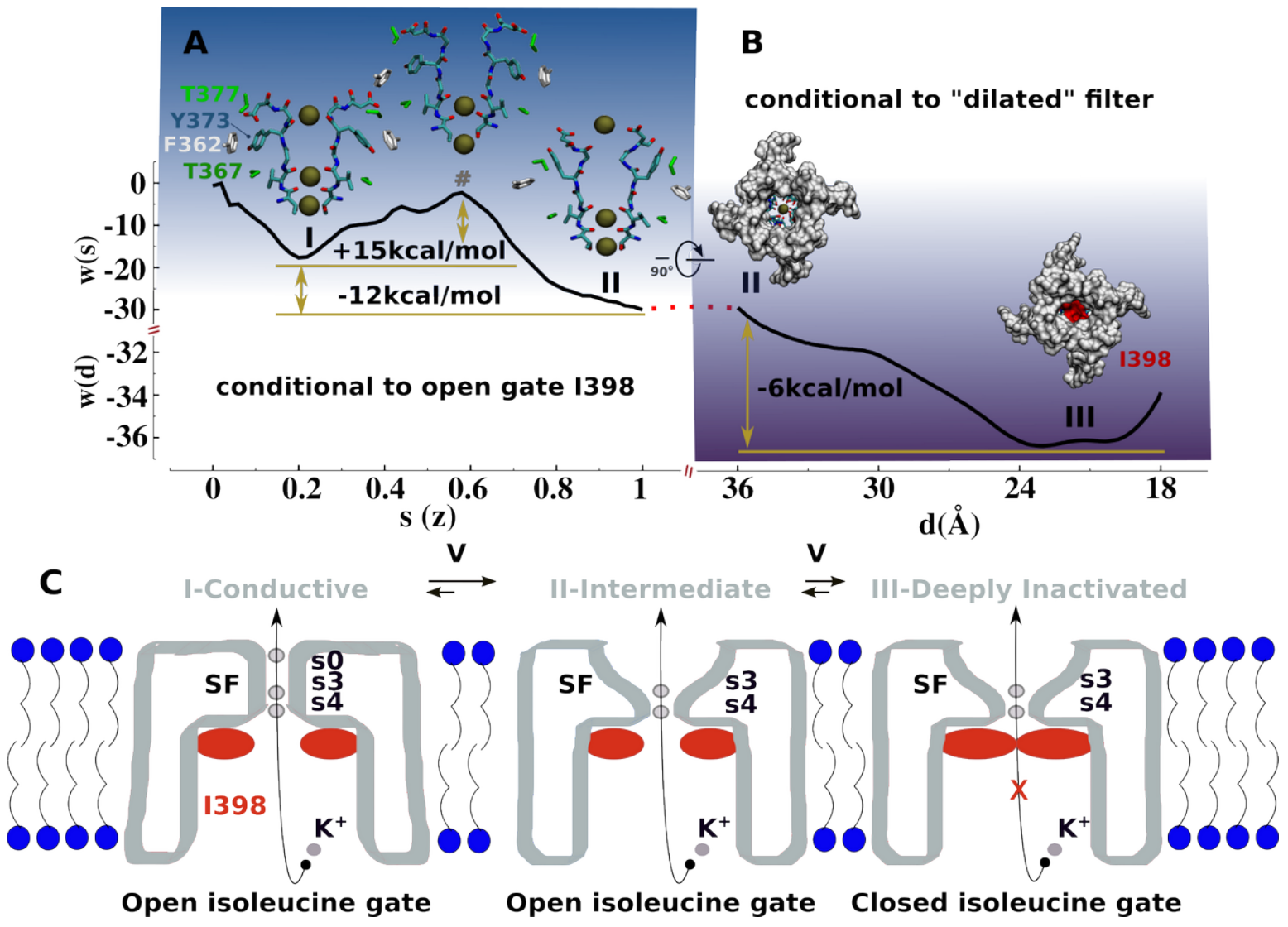

**Fig. 4. Mechanism of C-type inactivation of the triple-mutant channel kv1.2-kv2.1-3m.** (**A**) Free-energy profile w(s) along the conformational transition path s connecting the conductive (s=0.2) and dilated (s=1) states of the selectivity filter. The free-energy profile is conditional to the open configuration of the isoleucine gate. (**B**) Free-energy profile w(d) associated to closure of the isoleucine gate. The reaction coordinate d corresponds to the inter-subunit separation distance between ß-carbon atoms of I398. The free-energy profile is conditional to the dilated conformation of the selectivity filter. (**C**) C-type inactivation mechanism of the triple-mutant channel inferred from the representative structures of the selectivity filter and isoleucine gate along the free-energy profiles **A** and **B**.

Based on these findings, we conclude that the structures of kv1.2-kv2.1-3m (*9*), *Shaker*-W434F (*8*), Kv1.2-W366F (*10*) and *Shaker*-lowK (*12*) must correspond to a conductive metastable intermediate on the path toward the true non-conductive C-type inactivated state wherein the permeation pathway is blocked by the isoleucine gate. The free-energy landscape associated with the with C-type inactivation process comprises two sequential transitions between three metastable states: conductive → dilated-intermediate → deeply inactivated. Each transition was characterized by a separate potential of mean force(PMF) calculation (Methods, Fig. 4, Fig. S8 and S9). The first PMF monitoring the conformation of the selectivity filter while the isoleucine gate is in the open conformation shows two metastable states: conductive and dilated. The transition, which favors the latter state with a downhill free-energy change of -12 kcal/mol, involves a significant conformational change of Y373 from its buried configuration next to T367 in the conductive state towards its externally exposed orientation in the intermediate state resulting from unfavorable close interactions between the polar side chain of Y373 and F362. This is actually a metastable intermediate state that remains conductive. As shown by the second PMF monitoring the closure of the isoleucine gate toward a non-conductive deep inactivated state while the filter remains in dilated conformation shows a downhill change of -6.0 kcal/mol. Notwithstanding the limited accuracies of these computational estimates, the forward and backward free-energy barriers for these two transitions are in qualitative agreement with C-type inactivation of the triple-mutant channel in the sense that kv1.2-kv2.1-3m inactivates substantially faster than recovers from it—at +200mV, the time constant of C-type inactivation is predicted to be in the same microsecond range of the simulated deep inactivation of the channel (Fig. 3 and Fig. S3). The overall free-energy landscapes pointing towards the greater stability of the "deep inactivated" state is qualitatively correct.

The new structural insights into the intricate mechanism of C-type inactivation suggests new research directions in the field. Worth of investigation is the experimental observation that *Shaker* B leaks $Na^+$ in the absence or low concentration of $K^+$ in the C-type inactivation state (*35*). AMBER and CHARM36m simulations of kv1.2-kv2.1-3m in presence of 150mM NaCl show that one or two $Na^+$ ions can stably bind the dilated conformation of the selectivity filter at sites $S_3$, $S_4$ and $S_{ext}$, while favoring the open configuration of the isoleucine gate (Methods, Fig. S10 and Table S1). The internal $[Na^+]/[K^+]$ concentration ratio seems to affect the closing of the isoleucine gate with functional implications for sodium leak in C-type inactivation. Also important, it is the more extensive investigation of the structure-function relationship of the inactivation gate according to the amino-acid sequence within the pore domain. In particular, note that except for hERG, most studied potassium channels including, Kv1.2, *Shaker*-B, KcsA, BK and MthK, display either isoleucine, leucine, valine or phenylalanine at position 398 (Fig. S7). The hERG channel has a tyrosine at position 398 which in the cryo-EM structure (*36*), is twisted towards the central cavity in the open state. Because tyrosine is bulkier than asparagine, there is a chance that C-type inactivation in hERG also involves constriction of the permeation pathway at position 398 when the selectivity filter is dilated—another fascinating assumption worth of investigation in structural studies. Beyond its key role in C-type inactivation, the isoleucine gate may be also implicated in other molecular processes of Kv channels, including the recently discovered inactivation mechanism involving the electromechanical coupling between the S4S5 linker and the internal end of the pore of the Kv2.1 channel (*37*). As a piece of the molecular machinery implicated in multiple states, the isoleucine gate might interfere in a state-dependent manner with the mechanism of action of a broad class of QA blockers (*16*) and negatively charged activators that bind beneath the selectivity filter and operate as master keys to open a variety of $K^+$ channels (*38*). Of particular importance, the understanding of this last aspect could offer a distinct advantage in the development of drugs that modulate gating states of potassium channels more specifically.

## MATERIALS AND METHODS

**Molecular Dynamics.** The high-resolution x-ray structure of the triple-mutant Kv1.2 channel (kv1.2-kv2.1-3m) was obtained from the Protein Data Bank (PDB code 7SIT) (*9*). The channel structure was embedded in a fully hydrated (POPC) phospholipid bilayer and simulated at constant temperature 300K and pressure 1atm, neutral pH and with applied transmembrane (TM) electrostatic potential (Table S1). MD simulations were performed with Desmond using the purpose-built Anton2 supercomputer (*39*). Equations of motion were integrated using a time step of 2.5 fs and Van der Waals interactions were truncated at 12 Å. Ionic currents were driven by application of a constant electric field E across the simulation box to mimic a voltage clamp experiment at the depolarized voltage of +200 mV (corresponding to a TM electric field of 0.043 kcal/mol/Å/e) (*15*). Simulations were performed with three distinct all-atom force fields: AMBER, CHARMM36m and CHARMM36m-NBFIX. ff14SB version of the AMBER force field (*20*) was used in combination with ion parameters by Joungand and Cheatham (*21*), CHARMM36m (*22*) was used with standard ion parameters and CHARMM36m-NBFIX was used with modified ion parameters in which ion-carbonyl interactions were made more attractive to mimic the Joung and Cheatham model (Table S2). Water molecules were described by the TIP3P model (*40*). Setup and analysis of the MD trajectories was performed in VMD (*41*).

**Site-directed mutagenesis and RNA synthesis.** Kv1.2-kv2.1 chimera (kindly provided by Eduardo Perozo) was cloned into pMax vector flanked by *Xenopus* β-globin sequence. Mutagenesis was performed utilizing the QuickChange techniques. All the clones were verified with full length sequencing (Plasmidsaurus). DNA was linearized at the unique PmeI restriction site and then transcribed *in vitro* using T7 transcription kit (Ambion).

**Channel expression in *Xenopus* oocytes and electrophysiology.** Ovaries of *Xenopus laevis* were purchased from XENOPUS1. The follicular membrane was digested by collagenase type II (Worthington Biochemical Corporation) 2 mg/ml supplemented with bovine serum albumin (BSA) 1 mg/ml. Oocytes were incubated in standard oocytes solution (SOS) containing in mM: 96 NaCl, 2 KCl, 1.8 $CaCl_2$, 1 $MgCl_2$, 0.1 EDTA, 10 HEPES, and pH set to 7.4 with NaOH. SOS was supplemented with 50 mg/ml gentamycin. Stage V-VI oocytes were then selected and microinjected with 50 – 150ng of cRNA. Injected oocytes were maintained in SOS solution and kept at 18 °C for 1-4 days prior to recordings. Ionic currents were recorded using the cut-open voltage-clamp technique (*42*). Voltage-measuring pipettes were pulled using a horizontal puller (P-87 Model, Sutter Instruments, Novato, CA) with resistance between 0.3 and 0.8 MΩ were used to impale the oocytes. Currents were acquired by a setup comprising a Dagan CA-1B amplifier (Dagan, Minneapolis, MN) with a built-in low-pass four-pole Bessel filter for a cutoff frequency of 20 kHz. Using a 16-bit A/D converter (USB-1604, Measurement Computing, Norton, MA) for acquisition and controlled by an in- house software (GPatch64MC), data were sampled at 1 MHz, digitally filtered at Nyquist frequency and decimated for a storage acquisition rate of 100 kHz. Capacitive transient currents were compensated using a dedicated circuit. The voltage clamp was controlled by GPatch64MC and we used the USB-1604 16-bit as the D/A converter. Transient capacitive current was compensated by a dedicated circuit and in some cases, the transients were further minimized by an online P/N protocol holding at -80mV (*43*). All experiments were performed at room temperature (~17–18˚C) in external solution containing: (in mM) 120 potassium methylsulfonate (KMES), 2 calcium hydroxide, 0.1 EDTA and 10 HEPES, pH = 7.40 (with MES). Internal solution was composed by (in mM) 120 KMES, 10HEPES and 2EGTA, pH = 7.40 (with MES). Agitoxin II was obtained from Alomone Labs and was titrated to 100nM in the external solution prior to experiments. The current were elicited prior to the external application of the toxin. The blockage effects were assessed by series of 50 depolarizing pulses (from -120 to +60 mV) every 5 or 10s. Between experiments, 1% albumin solution (in water) was used to clean the chamber and the bridges. All chemicals used were purchased from Sigma-Aldrich (St. Louis, MO). GraphPad 9 (Prism), and in-house software (Analysis) were used to analyze the data.

**Binding Free-Energy of Charybdotoxin.** The high-resolution x-ray structures of kv1.2-kv2.1 (PDB code 2R9R) (*7*) and kv1.2-kv2.1-3m (PDB code 7SIT) (*9*) were used as molecular templates for modeling (*44*) the pore domain of the wild-type, double mutant (S367T/V377T) and triple mutant (W362F/S367T/V377T) constructs of the channel in the conductive and dilated conformational states, respectively. Binding of the molecular structure of charybdotoxin (CTX) (*45*) to each of the channel constructs was investigated with HDOCK (*46*), according to the condition that Lys27 of CTX is in close proximity to the external entrance of the selectivity filter. The root-mean square deviation between docking poses and the x-ray bound configuration of CTX (*18*) was considered as the structural criterion (rmsd ≤ 5Å) to select docking solutions best reproducing the bound state of the toxin. At least 10 independent docking solutions were selected for computation of the net free-energy difference ($\Delta\Delta G$) involved in the binding of CTX to each of the channel constructs and states.

$\Delta\Delta G$ was evaluated according to the continuous implicit solvent calculations of the PB-VDW model used in previous studies (*17*). The Poisson-Boltzmann (PB) solvation energy of the ligand-protein bound complex ($\Delta G_{LP}$) was calculated using the Adaptive Poisson-Boltzmann Solver 1.4.1 (APBS) (*47*) through a finite-difference scheme, by considering a 240 Å cubed box and a grid of 1.0 x 1.0 x 1.0 $Å^3$. By-representing explicitly the protein atoms without any charges, a dummy run was first carried out with APBS to generate dielectric, charge and accessibility maps for the molecule in solution. Following the molecular surface definition, the internal dielectric constant of the protein was set to 15. The electrolyte solution was represented with a dielectric constant of 80 and salt concentration of 100 mM. These maps were then modified for the inclusion of a low-dielectric ($\in = 2$) lipid surrogate. Input files and maps for APBS were generated with APBSmem (*48*). The van der Waals component of the bare electrostatic energy of the bound complex ($\Delta E_{LP}$) was computed with the CHARMM36m force field (*22*) by using the *namdenergy* plugin linked to VMD (*41*). The van der Waals component was scaled by an empirical factor ($\lambda$ = 0.17), intended to resolve the protein-solvent interaction, absent in the implicit solvent representation (*49*). Calculations included the two experimentally resolved bound potassium ions at sites $S_2/S_4$ and $S_3/S_4$ of the conductive and dilated conformation of the selectivity filter, respectively. Solvent accessible surface area (SASA) and entropic contributions (*50*) associated to the binding energy of CTX were assumed to be similar in both conformations of the channel and as such, they were not included in the calculation of the net binding free-energy difference $\Delta\Delta G$.

**Primary-Sequence Analysis.** Primary-sequence logos conservation throughout the main-pore segments PH, SF and S6 were generated with Weblogo3 (*51*) by taking into consideration a HMMER3.0 (*52*) generated multiple sequence alignment of 657 unique UniProt sequences.

**Energetics of C-type Inactivation.** The energetics of C-type inactivation was investigated by means of two potentials of mean force (PMFs). The first PMF reports the free-energy profile associated to the conformational transition of the selectivity filter between the conductive and dilated states, conditional to an open isoleucine gate. The second PMF reports the free-energy profile associated to closure of the isoleucine gate under the condition of a dilated conformation of the selectivity filter. Both conditions were imposed in the free-energy calculations via soft harmonic restraints of 0.5 kcal/mol/Å$^2$ respectively applied to α-carbon atoms of residue I398 and the selectivity filter. PMFs were determined employing the NAMD (*53*) implementation of the well-tempered metadynamics extended adaptive biasing force (WTM-eABF) algorithm (*54*, *55*), with the corrected *z*-averaged restraint (CZAR) estimator (*56*). Calculations were respectively carried out with CHARMM36m(*22*) and the ff14SB version of the AMBER force field (*20*).

The free-energy profile associated to the conformational transition of the selectivity filter between the conductive and dilated states was computed with two path-collective variable (PCVs) (*57*), formed by 22 internal atomic distances. The first PCV, *s*, corresponds to the path connecting the two end states of the transformation, *i.e.*, a string of discrete intermediate values inferred from an independent targeted molecular dynamics (TMD) simulation, whereas the second, orthogonal one, σ, represents the width of the tube embracing the path. The gradient of the free energy was measured along ***s***, while a soft harmonic potential with a force constant of 5 kcal/mol Å$^2$ was applied on w. No time-dependent bias was applied until a threshold of 50,000 samples was reached.

For the free-energy profile underlying the constriction of the pore domain, the collective variable (CV), $d=d_1+d_2$, was defined as the sum of two Euclidean distances separating the ß-carbon atom of residue I398 of subunits KCH1 and KCH3, on the one hand, and of subunits KCH2 and KCH4, on the other hand. The reaction pathway, $18 \leq d \leq 36$ Å, was discretized in bins 0.1 Å wide, wherein samples of the local force acting along the CV were accrued. To minimize nonequilibrium effects, no time-dependent bias was applied until a threshold of 10,000 samples was reached.

## ACKNOWLEDGMENTS

Helpful discussions with Eduardo Perozo and Leticia Stock are gratefully acknowledged. CHARMM36m-NBFIX parameters used in the study were gently provided by Ramon Mendoza Uriarte, member of Roux's lab. We thank Gethiely Gasparini for technical assistance with site-directed mutagenesis and RNA synthesis. Anton 2 computer time was provided by the Pittsburgh Supercomputing Center (PSC) through Grant MCB100018P from the National Institutes of Health. The Anton 2 machine at PSC was generously made available by D.E. Shaw Research. The work was supported by National Council for Scientific and Technological Development CNPq [WT grant number 302089/2019-5 and 200114/2020-4], by the National Institutes of Health Award R01GM030376, F.B. and National Science Foundation Award QuBBE QLCI (NSF OMA-2121044), F.B. B.P. is a PEW Latin American Fellow (2019).

## AUTHOR CONTRIBUTIONS

Conceptualization: WT and BR
Methodology: WT, YL, CB, BP, CC, FB and BR
Investigation: WT, YL, CB, BP, JA and CC
Visualization: WT, YL, JA and CC
Funding acquisition: WT, CC, FB and BR

Project Administration: WT and BR
Supervision: WT, FB, and BR
Writing (original draft): WT
Writing (review and editing): WT, YL, CB, BP, JA, CC, FB and BR

## COMPETING INTERESTS

The authors declare that they have no competing interests.

## DATA AND MATERIALS AVAILABILITY

All data needed to evaluate the conclusions of this work are included in the paper and/or the Supplementary Information.

## REFERENCES

1. T. Hoshi, W. N. Zagotta, R. W. Aldrich, Biophysical and molecular mechanisms of Shaker potassium channel inactivation. *Science* **250**, 533–538 (1990).

2. J. Ostmeyer, S. Chakrapani, A. C. Pan, E. Perozo, B. Roux, Recovery from slow inactivation in K+ channels is controlled by water molecules. *Nature* **501**, 121–124 (2013).

3. D. A. Doyle, J. M. Cabral, R. A. Pfuetzner, A. Kuo, J. M. Gulbis, S. L. Cohen, B. T. Chait, R. MacKinnon, The structure of the potassium channel: molecular basis of K+conduction and selectivity. *Science* **280**, 69–77 (1998).

4. Y. Zhou, J. H. Morais-Cabral, A. Kaufman, R. MacKinnon, Chemistry of ion coordination and hydration revealed by a K+channel-FAB complex at 2.0 Å resolution. *Nature* **414**, 43–48 (2001).

5. S. Bernèche, B. Roux, Energetics of ion conduction through the K+ channel. *Nature* **414**, 73–77 (2001).

6. W. Kopec, B. S. Rothberg, B. L. de Groot, Molecular mechanism of a potassium channel gating through activation gate-selectivity filter coupling. *Nat. Commun.* **10**, 5366 (2019).

7. S. B. Long, X. Tao, E. B. Campbell, R. MacKinnon, Atomic structure of a voltage-dependent K+ channel in a lipid membrane-like environment. *Nature* **450**, 376–382 (2007).

8. X.-F. Tan, C. Bae, R. Stix, A. I. Fernández-Mariño, K. Huffer, T.-H. Chang, J. Jiang, J. D. Faraldo-Gómez, K. J. Swartz, Structure of the Shaker Kv channel and mechanism of slow C-type inactivation. *Sci. Adv.* **8**, eabm7814 (2022).

9. R. Reddi, K. Matulef, E. A. Riederer, M. R. Whorton, F. I. Valiyaveetil, Structural basis for C-type inactivation in a Shaker family voltage-gated K+ channel. *Sci. Adv.* **8**, eabm8804 (2022).

10. W. Yangyu, Y. Yangyang, Y. Youshan, B. Shumin, R. Alberto, A. Ken, S. F. J, Cryo-EM structures of Kv1.2 potassium channels, conducting and non-conducting. *eLife* **12** (2023).

11. E. Perozo, R. MacKinnon, F. Bezanilla, E. Stefani, Gating currents from a nonconducting mutant reveal open-closed conformations in Shaker K+ channels. *Neuron* **11**, 353–358 (1993).

12. R. Stix, X.-F. Tan, C. Bae, A. I. Fernández-Mariño, K. J. Swartz, J. D. Faraldo-Gómez, Eukaryotic Kv channel Shaker inactivates through selectivity filter dilation rather than collapse. *Sci. Adv.* **9**, eadj5539 (2023).

13. P. Selvakumar, A. I. Fernández-Mariño, N. Khanra, C. He, A. J. Paquette, B. Wang, R. Huang, V. V. Smider, W. J. Rice, K. J. Swartz, J. R. Meyerson, Structures of the T cell potassium channel Kv1.3 with immunoglobulin modulators. *Nat. Commun.* **13**, 3854 (2022).

14. C. S. Souza, C. Amaral, W. Treptow, Electric fingerprint of voltage sensor domains. *Proc. Natl. Acad. Sci.* **111**, 17510–17515 (2014).

15. B. Roux, The membrane potential and its representation by a constant electric field in computer simulations. *Biophys. J.* **95**, 4205–4216 (2008).

16. G. Panyi, C. Deutsch, Probing the Cavity of the Slow Inactivated Conformation of Shaker Potassium Channels. *J. Gen. Physiol.* **129**, 403–418 (2007).

17. M. A. L. Eriksson, B. Roux, Modeling the structure of Agitoxin in complex with theShakerK+channel: a computational approach based on experimental distance restraints extracted from thermodynamic mutant cycles. *Biophys J* **83**, 2595–2609 (2002).

18. A. Banerjee, A. Lee, E. Campbell, R. MacKinnon, Structure of a pore-blocking toxin in complex with a eukaryotic voltage-dependent K+ channel. *eLife* **2** (2013).

19. L. G. Cuello, V. Jogini, D. Marien. Cortes, A. C. Pan, D. G. Gagnon, O. Dalmas, J. F. Cordero-Morales, S. Chakrapani, B. Roux, E. Perozo, Structural basis for the coupling between activation and inactivation gates in K+ channels. *Nature* **466**, 272–275 (2010).

20. J. A. Maier, C. Martinez, K. Kasavajhala, L. Wickstrom, K. E. Hauser, C. Simmerling, ff14SB: Improving the accuracy of protein side chain and backbone parameters from ff99SB. *J. Chem. Theory Comput.* **11**, 3696–3713 (2015).

21. I. S. Joung, T. E. I. Cheatham, Determination of Alkali and Halide Monovalent Ion Parameters for Use in Explicitly Solvated Biomolecular Simulations. *J. Phys. Chem. B* **112**, 9020–9041 (2008).

22. J. Huang, S. Rauscher, G. Nawrocki, T. Ran, M. Feig, B. L. de Groot, H. Grubmüller, A. D. MacKerell, CHARMM36m: An Improved Force Field for Folded and Intrinsically Disordered Proteins. *Nat. Methods* **14**, 71–73 (2017).

23. W. Treptow, M. Tarek, Molecular restraints in the permeation pathway of ion channels. *Biophys J* **91**, L26–L28 (2006).

24. M. Ø. Jensen, V. Jogini, M. P. Eastwood, D. E. Shaw, Atomic-level simulation of current–voltage relationships in single-file ion channels. *J. Gen. Physiol.*, doi: 10.1085/jgp.201210820 (2013).

25. Y. Yang, Y. Yan, F. J. Sigworth, How Does the W434F Mutation Block Current in Shaker Potassium Channels? *J. Gen. Physiol.* **109**, 779–789 (1997).

26. Z. Takacs, M. Toups, A. Kollewe, E. Johnson, L. G. Cuello, G. Driessens, M. Biancalana, A. Koide, C. G. Ponte, E. Perozo, T. F. Gajewski, G. Suarez-Kurtz, S. Koide, S. A. N. Goldstein, A designer ligand specific for Kv1.3 channels from a scorpion neurotoxin-based library. *Proc. Natl. Acad. Sci.* **106**, 22211–22216 (2009).

27. T. Kitaguchi, M. Sukhareva, K. J. Swartz, Stabilizing the Closed S6 Gate in the Shaker K v Channel Through Modification of a Hydrophobic Seal. *J. Gen. Physiol.* **124**, 319–332 (2004).

28. Y. Liu, C. A. Z. Bassetto, B. I. Pinto, F. Bezanilla, A mechanistic reinterpretation of fast inactivation in voltage-gated Na+ channels. *Nat. Commun.* **14**, 5072 (2023).

29. M. Holmgren, P. L. Smith, G. Yellen, Trapping of organic blockers by closing of voltage-dependent K+ channels - Evidence for a trap door mechanism of activation gating. *J Gen Physiol* **109**, 527–535 (NaN).

30. C. J. Peters, D. Fedida, E. A. Accili, Allosteric coupling of the inner activation gate to the outer pore of a potassium channel. *Sci. Rep.* **3**, 3025 (2013).

31. R. Olcese, D. Sigg, R. Latorre, F. Bezanilla, E. Stefani, A Conducting State with Properties of a Slow Inactivated State in a Shaker K+ Channel Mutant. *J. Gen. Physiol.* **117**, 149–164 (2001).

32. M. J. Lenaeus, M. Vamvouka, P. J. Focia, A. Gross, Structural basis of TEA blockade in a model potassium channel. *Nat. Struct. Mol. Biol.* **12**, 454–459 (2005).

33. L. G. Cuello, D. M. Cortes, E. Perozo, The gating cycle of a K+ channel at atomic resolution. *eLife* **6**, e28032 (2017).

34. A. J. Labro, D. M. Cortes, C. Tilegenova, L. G. Cuello, Inverted allosteric coupling between activation and inactivation gates in K+ channels. *Proc. Natl. Acad. Sci.* **115**, 5426–5431 (2018).

35. J. G. Starkus, L. Kuschel, M. D. Rayner, S. H. Heinemann, Ion conduction through C-type inactivated Shaker channels. *J. Gen. Physiol.* **110**, 539–550 (1997).

36. W. Wang, R. MacKinnon, Cryo-EM structure of the open human ether-à-go-go related K+ channel hERG. *Cell* **169**, 422-430.e10 (2017).

37. A. I. Fernández-Mariño, X.-F. Tan, C. Bae, K. Huffer, J. Jiang, K. J. Swartz, Inactivation of the Kv2.1 channel through electromechanical coupling. *Nature*, 1–8 (2023).

38. M. Schewe, H. Sun, Ü. Mert, A. Mackenzie, A. C. W. Pike, F. Schulz, C. Constantin, K. S. Vowinkel, L. J. Conrad, A. K. Kiper, W. Gonzalez, M. Musinszki, M. Tegtmeier, D. C. Pryde, H. Belabed, M. Nazare, B. L. de Groot, N. Decher, B. Fakler, E. P. Carpenter, S. J. Tucker, T. Baukrowitz, A pharmacological master key mechanism that unlocks the selectivity filter gate in K+ channels. *Science*, doi: 10.1126/science.aav0569 (2019).

39. D. E. Shaw, J. P. Grossman, J. A. Bank, B. Batson, J. A. Butts, J. C. Chao, M. M. Deneroff, R. O. Dror, A. Even, C. H. Fenton, A. Forte, J. Gagliardo, G. Gill, B. Greskamp, C. R. Ho, D. J. Ierardi, L. Iserovich, J. S. Kuskin, R. H. Larson, T. Layman, L.-S. Lee, A. K. Lerer, C. Li, D. Killebrew, K. M. Mackenzie, S. Y.-H. Mok, M. A. Moraes, R. Mueller, L. J. Nociolo, J. L. Peticolas, T. Quan, D. Ramot, J. K. Salmon, D. P. Scarpazza, U. B. Schafer, N. Siddique, C. W. Snyder, J. Spengler, P. T. P. Tang, M. Theobald, H. Toma, B. Towles, B. Vitale, S. C. Wang, C. Young, "Anton 2: Raising the Bar for Performance and Programmability in a Special-Purpose Molecular Dynamics Supercomputer" in *SC '14: Proceedings of the International Conference for High Performance Computing, Networking, Storage and Analysis* (2014), pp. 41–53.

40. W. L. Jorgensen, J. Chandrasekhar, J. D. Madura, R. W. Impey, M. L. Klein, Comparison of simple potential functions for simulating liquid water. *J. Chem. Phys.* **79**, 926–935 (1983).

41. W. Humphrey, A. Dalke, K. Schulten, VMD – Visual Molecular Dynamics. *J. Mol. Graph.* **14**, 33–38 (1996).

42. E. Stefani, F. Bezanilla, Cut-open oocyte voltage-clamp technique. *Methods Enzymol.* **293**, 300–318 (1998).

43. C. M. Armstrong, F. Bezanilla, Currents Related to Movement of the Gating Particles of the Sodium Channels. *Publ. Online 13 April 1973 Doi101038242459a0* **242**, 459–461 (1973).

44. B. Webb, A. Sali, Comparative Protein Structure Modeling Using MODELLER. *Curr. Protoc. Bioinforma.* **47**, 5.6.1-5.6.32 (2014).

45. F. Bontems, B. Gilquin, C. Roumestand, A. Ménez, F. Toma, Analysis of side-chain organization on a refined model of charybdotoxin: structural and functional implications. *Biochemistry* **31**, 7756–7764 (1992).

46. Y. Yan, H. Tao, J. He, S.-Y. Huang, The HDOCK server for integrated protein–protein docking. *Nat. Protoc.* **15**, 1829–1852 (2020).

47. N. A. Baker, D. Sept, S. Joseph, M. J. Holst, J. A. McCammon, Electrostatics of nanosystems: Application to microtubules and the ribosome. *Proc. Natl. Acad. Sci.* **98**, 10037–10041 (2001).

48. K. M. Callenberg, O. P. Choudhary, G. L. de Forest, D. W. Gohara, N. A. Baker, M. Grabe, APBSmem: A Graphical Interface for Electrostatic Calculations at the Membrane. *PLoS ONE* **5**, e12722 (2010).

49. P. Nandigrami, F. Szczepaniak, C. T. Boughter, F. Dehez, C. Chipot, B. Roux, Computational Assessment of Protein–Protein Binding Specificity within a Family of Synaptic Surface Receptors. *J. Phys. Chem. B* **126**, 7510–7527 (2022).

50. M. K. Gilson, J. A. Given, B. L. Bush, J. A. McCammon, The statistical-thermodynamic basis for computation of

binding affinities: a critical review. *Biophys. J.* **72**, 1047–1069 (1997).

51. G. E. Crooks, G. Hon, J.-M. Chandonia, S. E. Brenner, WebLogo: a sequence logo generator. *Genome Res.* **14**, 1188–1190 (2004).

52. R. D. Finn, J. Clements, S. R. Eddy, HMMER web server: interactive sequence similarity searching. *Nucleic Acids Res.* **39**, W29–W37 (2011).

53. J. C. Phillips, D. J. Hardy, J. D. C. Maia, J. E. Stone, J. V. Ribeiro, R. C. Bernardi, R. Buch, G. Fiorin, J. Hénin, W. Jiang, R. McGreevy, M. C. R. Melo, B. K. Radak, R. D. Skeel, A. Singharoy, Y. Wang, B. Roux, A. Aksimentiev, Z. Luthey-Schulten, L. V. Kalé, K. Schulten, C. Chipot, E. Tajkhorshid, Scalable molecular dynamics on CPU and GPU architectures with NAMD. *J. Chem. Phys.* **153**, 044130 (2020).

54. H. Fu, H. Zhang, H. Chen, X. Shao, C. Chipot, W. Cai, Zooming across the Free-Energy Landscape: Shaving Barriers, and Flooding Valleys. *J. Phys. Chem. Lett.* **9**, 4738–4745 (2018).

55. H. Fu, X. Shao, W. Cai, C. Chipot, Taming Rugged Free Energy Landscapes Using an Average Force. *Acc. Chem. Res.* **52**, 3254–3264 (2019).

56. A. Lesage, T. Lelièvre, G. Stoltz, J. Hénin, Smoothed Biasing Forces Yield Unbiased Free Energies with the Extended-System Adaptive Biasing Force Method. *J. Phys. Chem. B* **121**, 3676–3685 (2017).

57. D. Branduardi, F. L. Gervasio, M. Parrinello, From A to B in free energy space. *J. Chem. Phys.* **126**, 054103 (2007).

# Supporting Information for

## Isoleucine gate blocks $K^+$ conduction in C-type inactivation

Werner Treptow *et al.*

*Corresponding author. Email: treptow@unb.br, roux@uchicago.edu

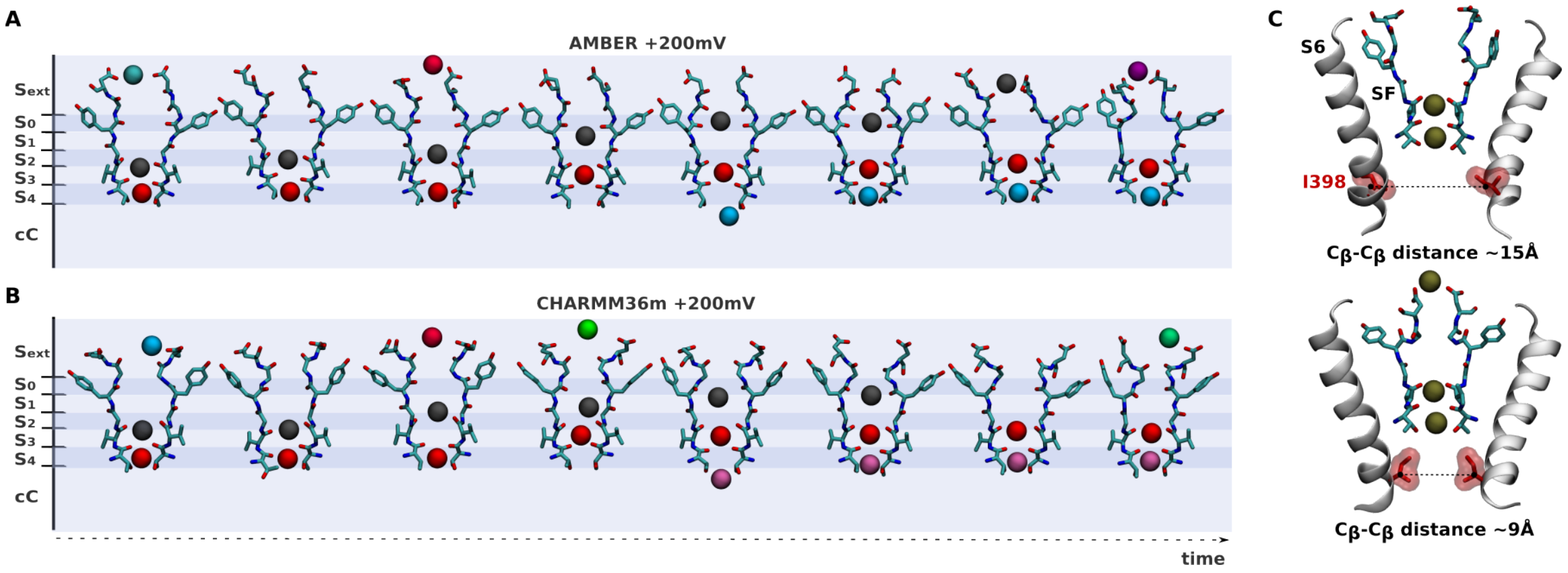

**Fig. S1. Ion conduction across the dilated conformation of the selectivity filter of kv1.2-kv2.1-3m. (A, B)** Shown are representative single-ion conduction events across the selectivity filter of the triple-mutant channel along AMBER and CHARMM36m simulations. (**C**) Closure of the isoleucine gate blocks ion conduction across the the selectivity filter.

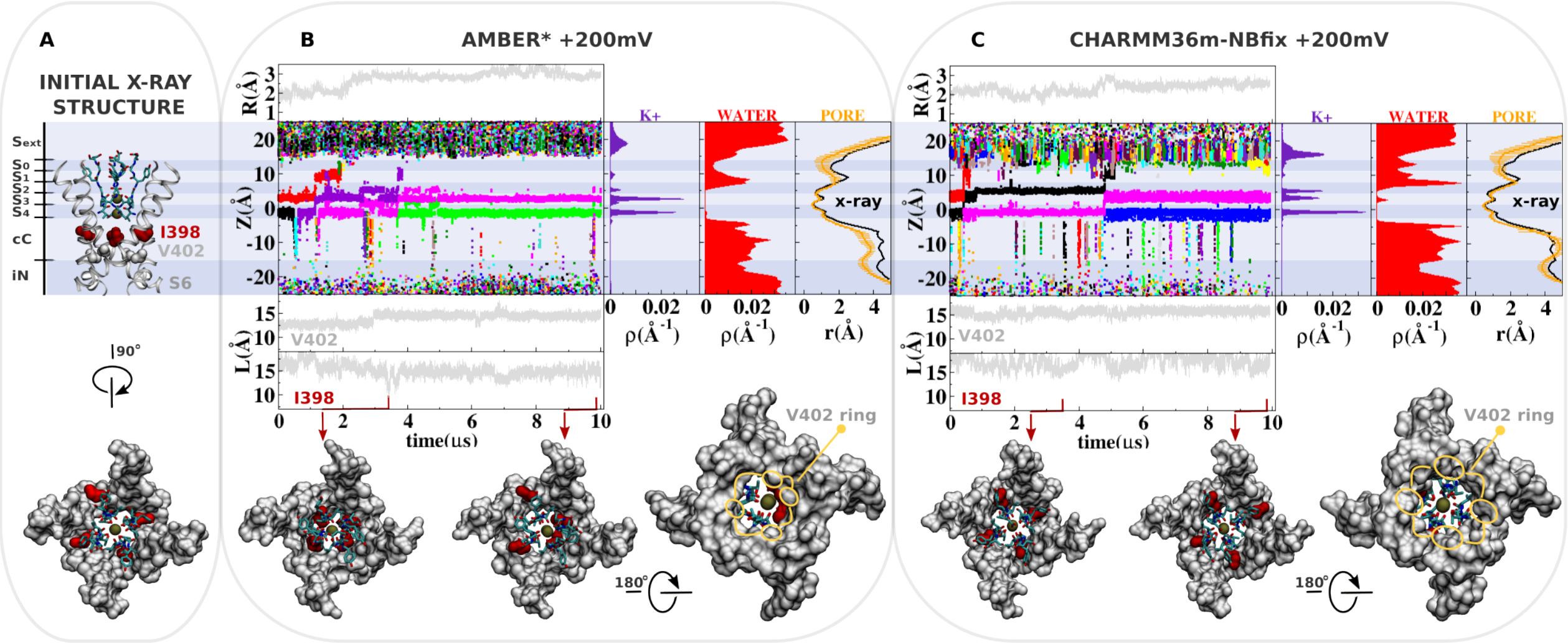

**Fig. S2. MD simulation of kv1.2-kv2.1-3m at +200mV.** (**A**) Molecular representation of the main pore of the channel, highlighting the initial configuration of the selectivity filter, I398 (red) and V402 (light gray). (**B** and **C**) Analysis of AMBER* and CHARMM36m-NBFIX trajectories. Shown is the structural deviation of the selectivity filter (R), trajectory of $K^+$ along the permeation pathway (Z) and the intersubunit $C_\beta$-$C_\beta$ separation distance of I398 and V402 (L) as a function of simulation time. Inset shows instantaneous configurations of I398 (red arrows). Time averages are the linear density of $K^+$ ions, the linear density of waters and the pore radius profile (r) along the permeation pathway Z.

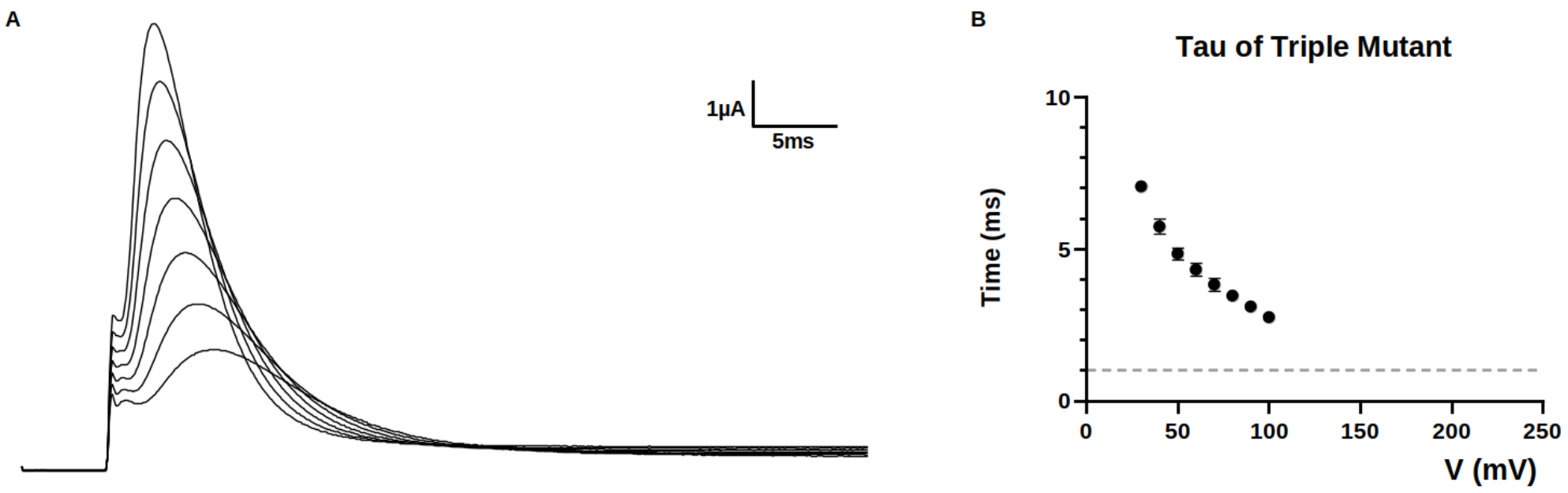

**Fig. S3. Decay of the ionic current of the triple-mutant kv1.2-kv2.1-3m channel.** (**A**) Ionic traces from the triple mutant channel and (**B**) its time constants fitted with a one exponential decay. When extrapolated to +200mV where the simulation was performed, the time constant (tau) was in the microsecond range (gray dashed line indicates 1ms), suggesting the time scale of the simulation and the conformational changes seen were relevant in physiological terms. Data plotted with Mean SEM.

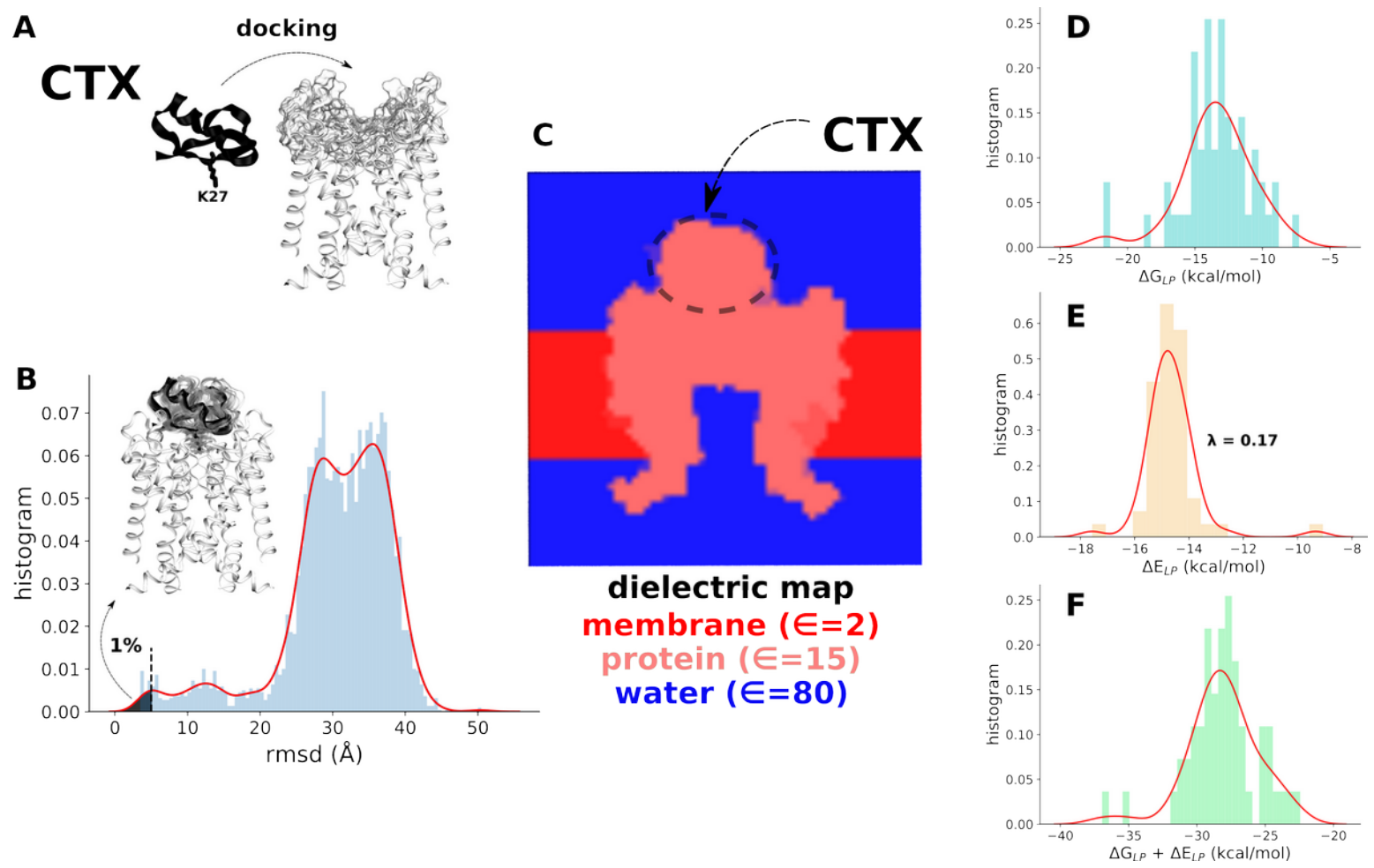

**Fig. S4. Estimation of the binding free-energy of CTX.** (**A**) Docking of CTX to the selectivity filter of the kv1.2-kv2.1 chimera channel. (**B**) Docking solutions (light black traces) best reproducing the experimentally resolved bound state of the toxin (tick black trace, PDB 4JTD). (**C**) Dielectric (∈) map of the toxin-protein complex considered in the Poisson-Boltzmann calculation. (**D**, **E** and **F**) Energy distribution of best docking solutions shown in (**B**). Poisson-Boltzmann ($\Delta G_{LP}$) and van der Waals ($\Delta E_{LP}$) electrostatic contributions to the binding free-energy of CTX are shown. The van der Waals component was scaled by an empirical factor λ = 0.17, intended to resolve the protein-solvent interaction absent in the implicit solvent representation. The same procedure was adopted to estimate the binding energy of CTX to all channel constructs and states (*cf.* Table S3).

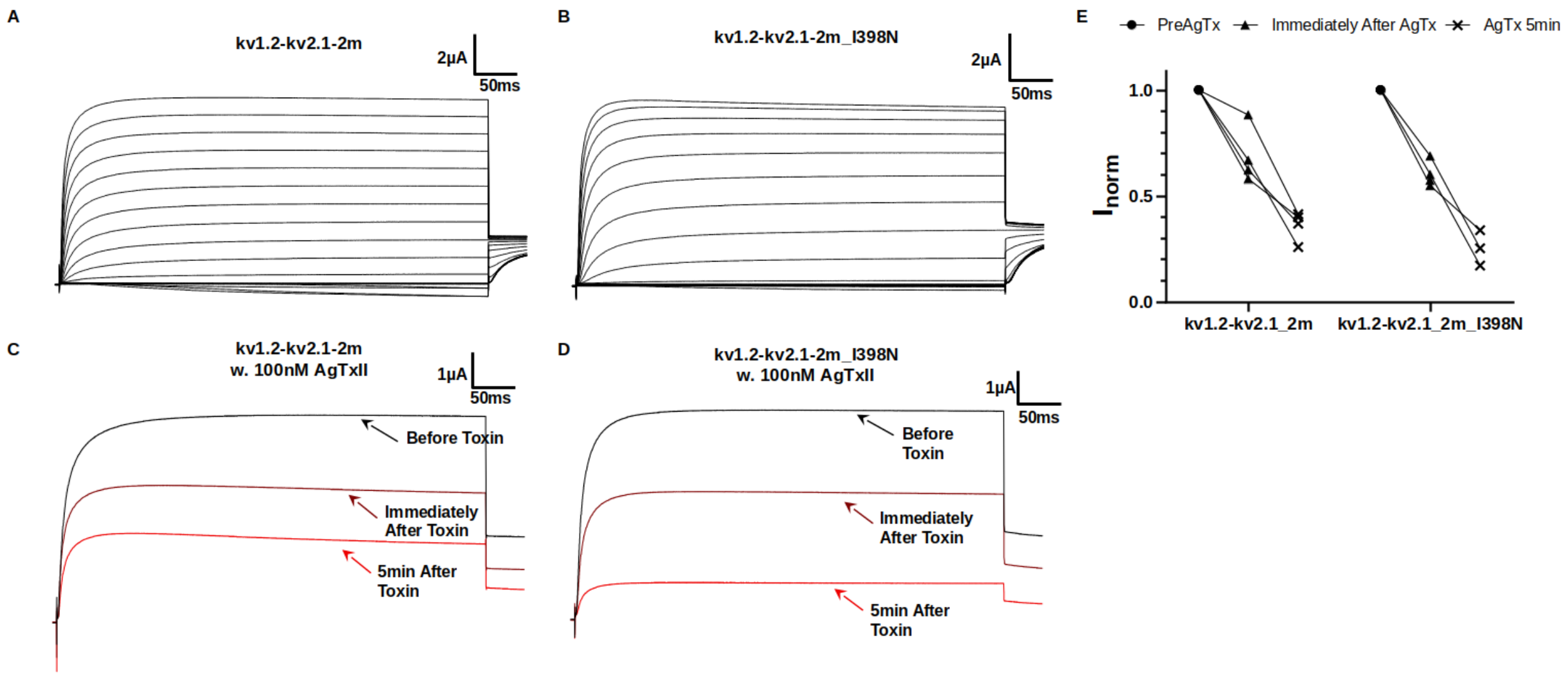

**Fig. S5. State dependence of I398N effects**. (**A**) Macroscopic current from the double mutant chimera channel kv1.2-kv2.1-2m (S367T/V377T). (**B**) Macroscopic current from kv1.2-kv2.1-2m with I398N substitution. Mutation I398N in the absence of W362F does not significantly alter the phenotype of the channel. (**C**, **D** and **E**) Effects of 100nM AgitoxinII (AgTxII) on the double mutant kv1.2-kv2.1-2m and double mutant kv1.2-kv2.1-2m with I398N substitution. AgTxII binds and blocks efficiently both channel constructs, suggesting the effect of I398N depends on the C-type inactivation induced by the W362F mutation.

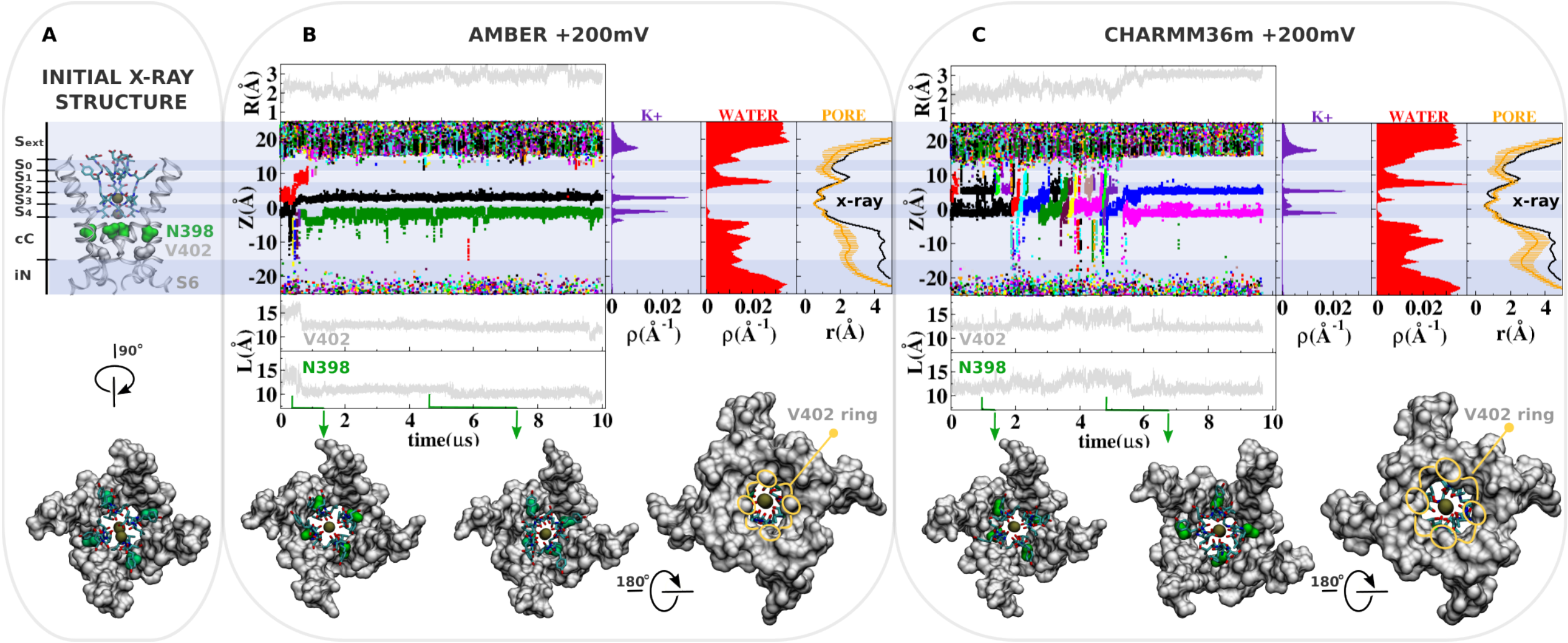

**Fig. S6. MD simulation of kv1.2-kv2.1-3m with I398N at +200mV.** (**A**) Molecular representation of the main pore of the channel, highlighting the initial configuration of the selectivity filter, N398 (green) and V402 (light gray). (**B** and **C**) Analysis of AMBER and CHARMM36m trajectories. Shown is the structural deviation of the selectivity filter (R), trajectory of $K^+$ along the permeation pathway (Z) and the intersubunit $C_\beta$-$C_\beta$ separation distance of N398 and V402 (L) as a function of simulation time. Inset shows instantaneous configurations of N398 (green arrows). Time averages are the linear density of $K^+$ ions, the linear density of waters and the pore radius profile (r) along the permeation pathway Z.

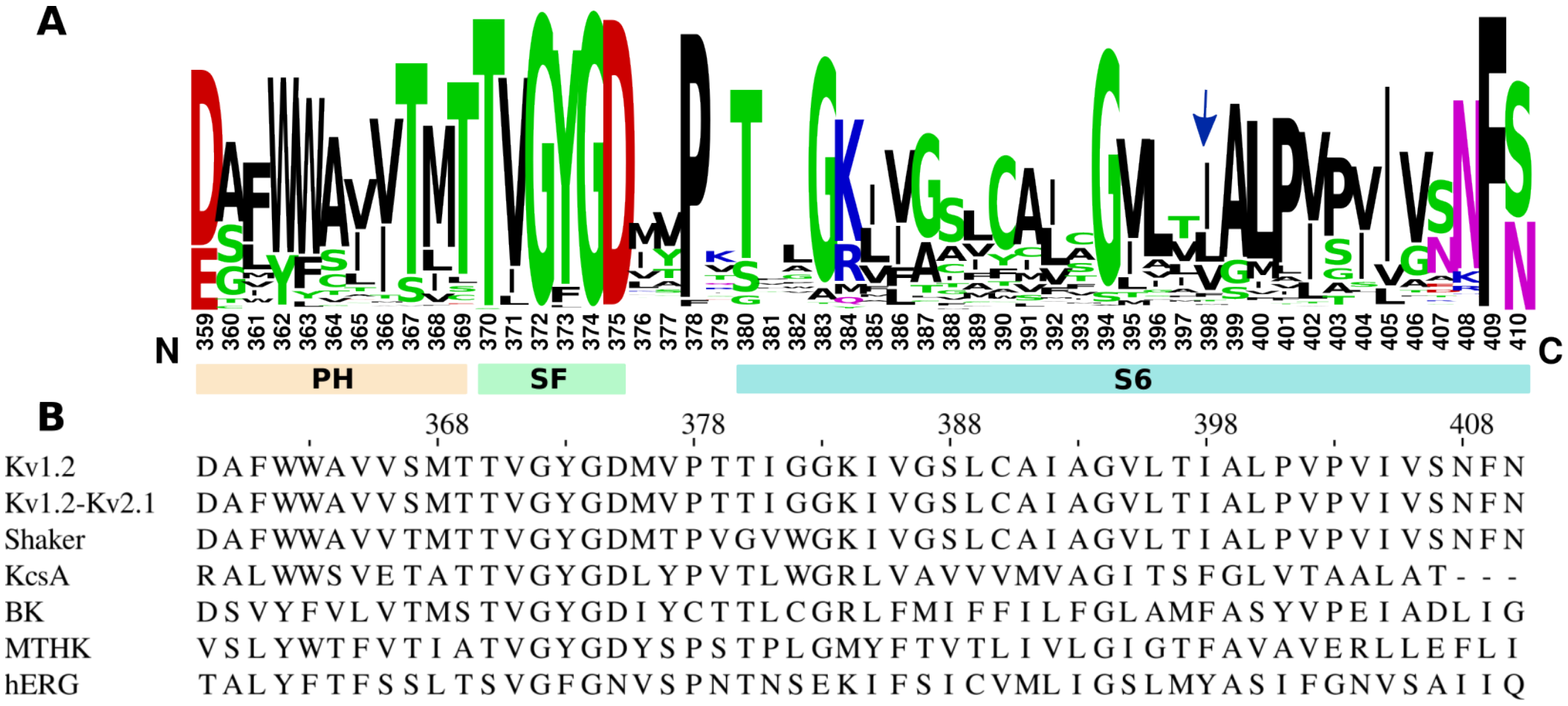

**Fig. S7. Primary sequence conservation throughout the main-pore segments PH, SF and S6.** (**A**) Logos conservation across $K^+$ channels. (**B**) Multiple-sequence alignment of most studied $K^+$ channels.

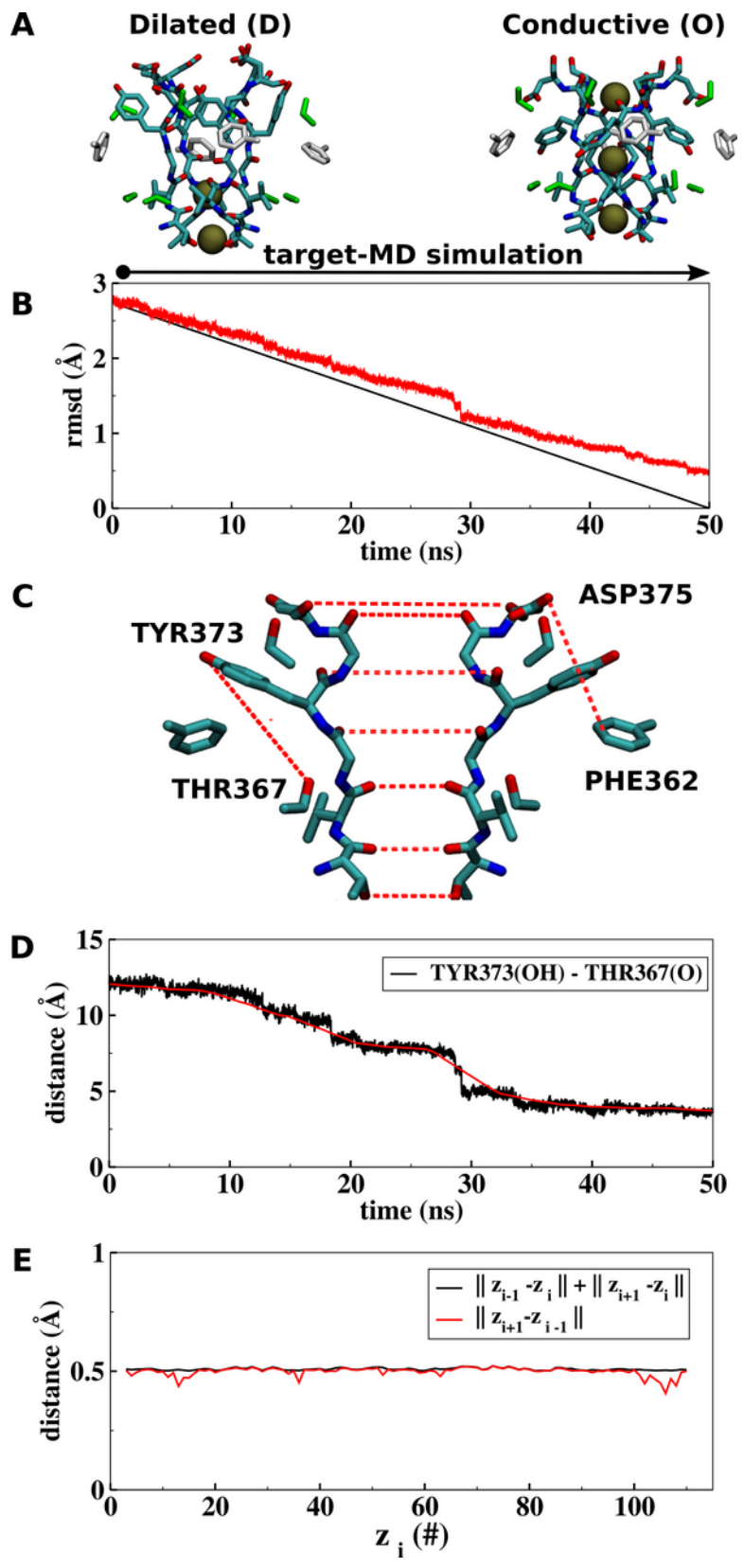

**Fig. S8. Conformational path between the conductive (O) and dilated (D) states of the selectivity filter.** (**A**, **B**) Time evolution of the instantaneous (red) and target (black) root-mean-square deviation (rmsd) of the selectivity filter between states O and D. Reference structures of the dilated and conductive states of the selectivity filter were respectively defined on basis of the high-resolution x-ray structures of the triple-mutant kv1.2-kv2.1-3m and kv1.2-kv2.1 (PDB codes 7SIT and 2R9R).TMD was carried out for 50 ns with an applied constant force of 400 kcal/mol/Å$^2$, corresponding to a *per* atom constant force of 2.38 kcal/mol/Å$^2$. (**C**) A total of 22 atomic distances were used for definition of the conformational path between states D and O (red dashes): 14 main-chain carbonyl distances between two opposing subunits of the channel, 4 side-chain distances between TYR373(OH) and THR367(O) and, 4 side-chain distances between ASP375(OD2) and PHE362(CD1). Atomic distances were symmetrized across the channel to support convergence of the calculation. (**D**) Time evolution of a representative coordinate of the transformation path. The coordinate was smoothed out following a running average procedure of the data. The running average length was 10% of the total number of data points *i.e.*, 500. (red). (E) The resulting string **z** consisted of 111 equidistant points i satisfying the monotonic condition $\|z_i - z_{i-1}\| + \|z_{i+1} - z_i\| \le \|z_{i+1} - z_{i-1}\|$ at a distance interval of 0.25 Å.

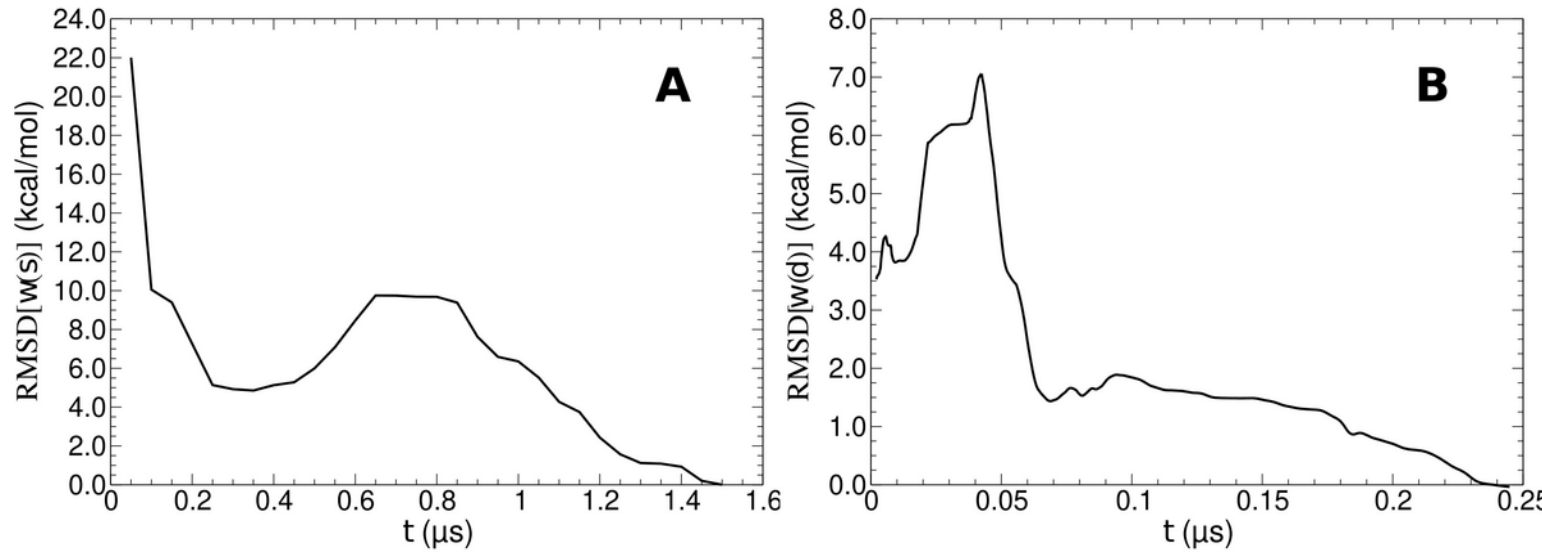

**Fig. S9. Convergence analysis of free-energy calculations.** (**A**, **B**) Respectively shown is the root-mean-square deviation (RMSD) of the free-energy profiles w(s) and w(d) as a function of simulation time t.

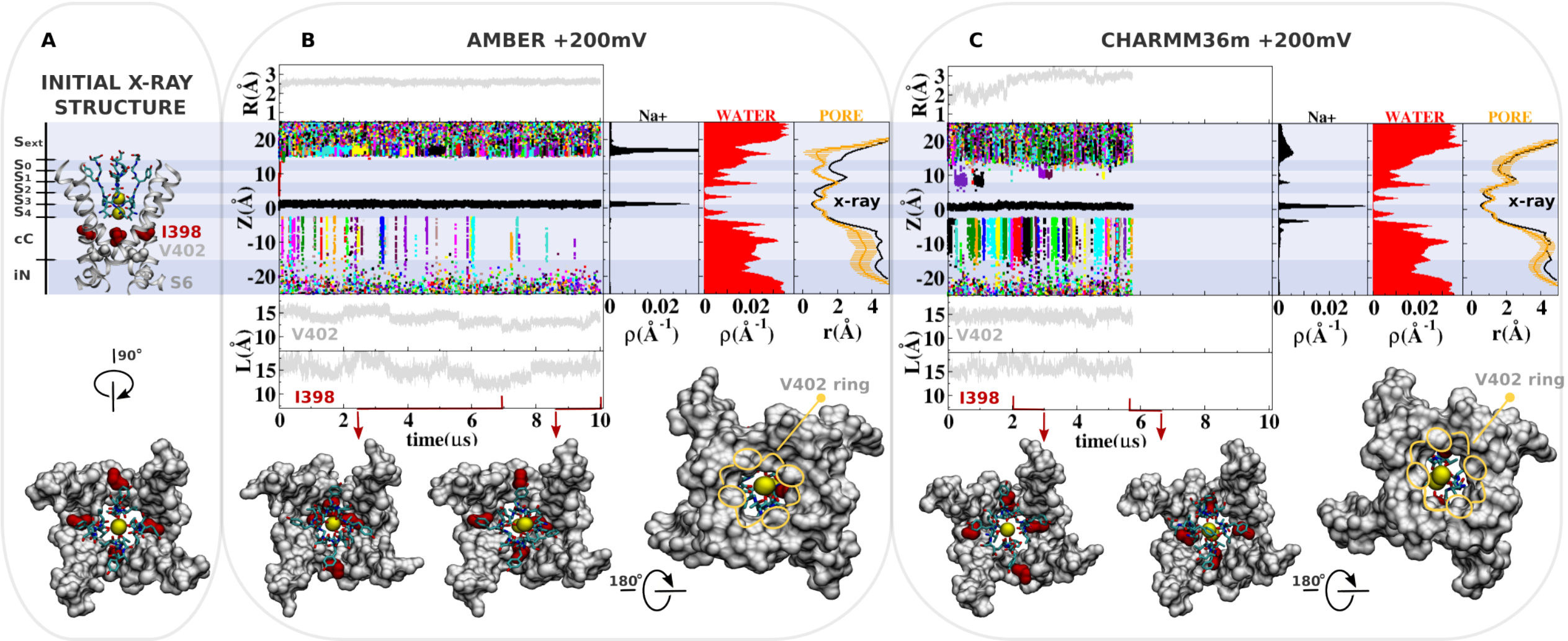

**Fig. S10. MD simulation of kv1.2-kv2.1-3m at +200mV and 150mM NaCl.** (**A**) Molecular representation of the main pore of the channel, highlighting the initial configuration of the selectivity filter, I398 (red) and V402 (light gray). Two sodium ($Na^+$) ions (yellow) are shown in the selectivity filter at the ion binding sites $S_3$ and $S_4$. (**B** and **C**) Analysis of AMBER and CHARMM36m trajectories. Shown is the structural deviation of the selectivity filter (R), trajectory of $Na^+$ along the permeation pathway (Z) and the intersubunit $C_\beta$-$C_\beta$ separation distance of I398 and V402 (L) as a function of simulation time. Inset shows instantaneous configurations of I398 (red arrows). Time averages are the linear density of $Na^+$ ions, the linear density of waters and the pore radius profile (r) along the permeation pathway Z.

## Table S1. MD simulations of triple mutant channel kv1.2-kv2.1-3m

| Simulation | Force Field | Voltage (mV) | Solution (mM) | Mutation | Restraint (kcal/mol)* | Time scale (μs) |
|---|---|---|---|---|---|---|
| **1** | AMBER | +200 | 150 mM KCl | W362F, S367T,V377T | 0 | 10.0 |
| **2** | AMBER* | +200 | 150 mM KCl | W362F, S367T,V377T | 1 | 10.0 |
| **3** | AMBER | +200 | 150 mM KCl | W362F, S367T,V377T,I398N | 0 | 10.0 |
| **4** | CHARMM36m | +200 | 150 mM KCl | W362F, S367T,V377T | 0 | 10.0 |
| **5** | CHARMM36m | +200 | 150 mM KCl | W362F, S367T,V377T,I398N | 0 | 10.0 |
| **6** | CHARMM36m-NBFIX | +200 | 150 mM KCl | W362F, S367T,V377T | 0 | 10.0 |
| **7** | AMBER | +200 | 150 mM NaCl | W362F, S367T,V377T | 0 | 10.0 |
| **8** | CHARMM36m | +200 | 150 mM NaCl | W362F, S367T,V377T | 0 | 10.0 |

*Harmonic restraints were applied to $C_\beta$ atoms of I398 to keep the isoleucine gate in the open state

## Table S2. NBFixes for potassium, carbonyl and water interactions

| Interactions | Atom Type1 | Atom Type 2 | $E_{min}$ (Kcal/mol) | $R_{min}$ (Å) |
|---|---|---|---|---|
| Potassium-Water | POT | OT | -0.015033 | 4.211 |
| Potassium-Carbonyl | POT | O | -0.455556 | 3.044 |
| Water-Carbonyl | OT | O | -0.338333 | 3.200 |

## Table S3. Number of conduction events along MD simulations in which the isoleucine gate is open

| Simulation | Force Field | Voltage (mV) | Conduction events (#) | Time scale (μs) | Single-channel conductance (pS) |
|---|---|---|---|---|---|
| **2** | AMBER* | +200 | 2 | 10.0 | 0.16 |
| **4** | CHARMM36m | +200 | 40 | 10.0 | 3.20 |
| **6** | CHARMM36m-NBFIX | +200 | 2 | 10.0 | 0.16 |
| **Combined number of conduction events** | | | 44 | 30.0 | 1.17 |

# Table S4. Binding free-energy difference of CTX

| Channel | Mutations | $\triangle G_{LP}^1$ (kcal/mol) | $\triangle E_{LP}^1$ (kcal/mol) | $\triangle G_{LP}^2$ (kcal/mol) | $\triangle E_{LP}^2$ (kcal/mol) | $\triangle\triangle G^*$ (kcal/mol) |
|---|---|---|---|---|---|---|
| kv1.2-kv2.1 | wild-type | -13.43 ± 2.67 | -14.65 ± 1.01 | 6.77 ± 3.93 | -13.93 ± 1.43 | -20.93 ± 4.44 |
| kv1.2-kv2.1-2m | S367T/V377T | -4.06 ± 4.09 | -14.66 ± 1.20 | 4.72 ± 5.58 | -13.41 ± 1.20 | -10.03 ± 6.45 |
| kv1.2-kv2.1-3m | W362F/S367T/V377T | -5.02 ± 3.55 | -15.28 ± 1.32 | 6.25 ± 5.61 | -14.66 ± 1.64 | -11.90 ± 5.77 |

$^*\triangle\triangle G = (\triangle G_{LP}^1+\triangle E_{LP}^1) - (\triangle G_{LP}^2+\triangle E_{LP}^2)$ is the net free-energy difference of CTX binding to the conductive (1) and dilated (2) conformations of the pore domain. Each free-energy estimate and statistical error was determined based on at least 10 independent docking solutions best describing the x-ray bound state of CTX.